\documentclass[twocolumn]{aastex701}

\usepackage{amsmath}
\begin{document}

\title{SPHEREx Ultracool Dwarf spectrophotometric Atlas (SUDA): Atmospheric and Fundamental Parameters of Ultracool Dwarfs}

\correspondingauthor{Shu Wang}
\email{shuwang@nao.cas.cn}

\author[0009-0000-7976-7383]{Zhijun Tu}
\affiliation{CAS Key Laboratory of Optical Astronomy, National Astronomical Observatories, Chinese Academy of Sciences, Beijing 100101, People's Republic of China}
\email{zjtu@bao.ac.cn}

\author[0000-0003-4489-9794]{Shu Wang}
\affiliation{CAS Key Laboratory of Optical Astronomy, National Astronomical Observatories, Chinese Academy of Sciences, Beijing 100101, People's Republic of China}
\affiliation{School of Astronomy and Space Sciences, University of Chinese Academy of Sciences, Beijing 100049, People's Republic of China}
\email{shuwang@nao.cas.cn}

\author[0009-0004-3407-0848]{Haomiao Huang}
\affiliation{CAS Key Laboratory of Optical Astronomy, National Astronomical Observatories, Chinese Academy of Sciences, Beijing 100101, People's Republic of China}
\affiliation{School of Astronomy and Space Sciences, University of Chinese Academy of Sciences, Beijing 100049, People's Republic of China}
\email{huanghm@bao.ac.cn}

\author[0000-0001-7084-0484]{Xiaodian Chen}
\affiliation{CAS Key Laboratory of Optical Astronomy, National Astronomical Observatories, Chinese Academy of Sciences, Beijing 100101, People's Republic of China}
\affiliation{School of Astronomy and Space Sciences, University of Chinese Academy of Sciences, Beijing 100049, People's Republic of China}
\affiliation{Institute for Frontiers in Astronomy and Astrophysics, Beijing Normal University, Beijing 102206, People's Republic of China}
\email{chenxiaodian@nao.cas.cn}

\author[0000-0002-2874-2706]{Jifeng Liu}
\affiliation{CAS Key Laboratory of Optical Astronomy, National Astronomical Observatories, Chinese Academy of Sciences, Beijing 100101, People's Republic of China}
\affiliation{School of Astronomy and Space Sciences, University of Chinese Academy of Sciences, Beijing 100049, People's Republic of China}
\affiliation{Institute for Frontiers in Astronomy and Astrophysics, Beijing Normal University, Beijing 102206, People's Republic of China}
\affiliation{New Cornerstone Science Laboratory, National Astronomical Observatories, Chinese Academy of Sciences, Beijing 100012, People's Republic of China}
\email{jfliu@nao.cas.cn}

\begin{abstract}

We present the SPHEREx Ultracool Dwarf spectrophotometric Atlas (SUDA), a homogeneous sample of 1675 field ultracool dwarfs with continuous 0.75--5~$\mu$m spectrophotometry from SPHEREx Quick Release 2 (QR2). Using the SAND and ATMO2020++ atmospheric grids, we derive atmospheric parameters, compute bolometric luminosities ($L_{\rm bol}$), and combine $T_{\mathrm{eff}}$ and radii with evolutionary tracks to estimate masses, ages, and evolutionary $\log g$. For sources at 1700--2500~K, spectroscopic $\log g$ is systematically lower than evolutionary $\log g$, with a median offset of $\sim$1.05~dex, likely driven primarily by cloud-related model uncertainties. We further construct an empirical atlas by binning the measurements in $T_{\mathrm{eff}}$ and evolutionary $\log g$, producing 57 spectrophotometric templates spanning $T_{\mathrm{eff}}\simeq700$--3000~K. Molecular indices trace a coherent atmospheric sequence: H$_2$O and CH$_4$ indices strengthen toward lower $T_{\mathrm{eff}}$, while CO and CO$_2$ indices increase below $\sim$1500~K and turn over near $\sim$1000~K. Model comparisons indicate that CO$_2$ index is more sensitive to metallicity than gravity at $T_{\mathrm{eff}}\sim800$--1200~K. SUDA provides a reference sample linking 0.75--5~$\mu$m spectrophotometric morphology to atmospheric and evolutionary trends in ultracool dwarfs.

\end{abstract}


\section{Introduction} \label{sec:intro}

Ultracool dwarfs, spanning the latest M dwarfs through the L, T, and Y spectral classes, occupy the coolest and lowest mass end of stellar and substellar populations \citep{Kirkpatrick1999L,Burgasser2002T,Cushing2011Y}. Their atmospheric parameters, including effective temperature ($T_{\rm eff}$), surface gravity ($\log g$), and metallicity ([M/H]), together with derived physical properties such as bolometric luminosity ($L_{\rm bol}$), mass, radius, and age, are central to understanding star formation at the lowest masses, the boundary between stars and brown dwarfs, and the connection between stars, brown dwarfs, and giant exoplanets. These quantities are also relevant to the radiation environments and potential habitability of planets around low mass stars, since stellar activity and ultraviolet emission in M dwarfs evolve with spectral type and age and strongly affect circumstellar habitable zone conditions \citep{2013MNRAS.431.2063S,2014AJ....148...64S,2024ApJ...966...69L,2025ApJS..281...13L}. Because ultracool dwarfs emit most of their energy in the near- and mid-infrared, infrared spectroscopy is essential for reliable physical characterization \citep{2025ApJ...979..145L,2025ApJ...994..237M,2025ApJ...982L..38V,2026A&A...707A..92L}.

Among these quantities, the bolometric luminosity plays a central and unifying role. Unlike atmospheric parameters inferred from spectral fitting, $L_{\rm bol}$ is comparatively independent of atmospheric models and links spectroscopy to derived physical properties through evolutionary models \citep{2015A&ABaraffe,2020A&A...637A..38P,2023A&AChabrier}. Accurate determinations of $L_{\rm bol}$ are critical for constraining masses, radii, and ages, and for breaking degeneracies among $T_{\rm eff}$, $\log g$, and metallicity that commonly arise in atmospheric fitting \citep{2023ApJSanghi,2024ApJ...973..107B}. Improving the accuracy and homogeneity of bolometric luminosity measurements is therefore essential for a more physically consistent view of the ultracool dwarf population.

One persistent limitation has been the incomplete spectral coverage. As ultracool atmospheres cool, an increasing fraction of the emergent flux shifts beyond $\sim$2.5~$\mu$m, making the 3--5~$\mu$m region critical for both luminosity estimates and atmospheric interpretation \citep{SaumonMarley2008,Marley2021Sonora}. This wavelength range also contains key molecular diagnostics, especially CO and CO$_2$, that are sensitive to chemistry, composition, and thermal structure in the coldest brown dwarfs \citep{2024AJ....167..237L,2024AJ....167..168M,2025Natur.645...62F,2026arXiv260604294L}.

Previous large spectroscopic studies of ultracool dwarfs focused on optical and near infrared wavelengths shortward of $\sim$2.5~$\mu$m, with widely used libraries built primarily from facilities such as IRTF/SpeX and Magellan/FIRE \citep{2021ApJ...921...95Z,2025ApJ...982...79B,2026AJ....171..198M}. As a result, many luminosity estimates still rely on heterogeneous SED assembly, combining partial spectra with sparse mid infrared constraints and model interpolation or extrapolation across unobserved regions \citep{2014AJDieterich,2015ApJFilippazzo,2023ApJSanghi}. For example, \citet{2023ApJSanghi} derived atmospheric parameters and physical properties for more than a thousand ultracool dwarfs from optical to mid infrared SED compilations and atmosphere model comparisons. That work established an important benchmark, but the spectroscopic coverage was limited to 1--2.5~$\mu$m, with the remainder of the SED constrained only by photometry.

The SPHEREx (Spectro-Photometer for the History of the Universe, Epoch of Reionization and Ices Explorer) mission provides a direct way to overcome this limitation through all sky spectrophotometry from 0.75--5~$\mu$m with continuous wavelength coverage \citep{spherex2018,spherex2020,spherex2025,spherexpipeline,2025IRSA652}. Its combination of full sky coverage and continuous infrared spectrophotometry enables homogeneous characterization of a vast ultracool dwarf population. In particular, SPHEREx provides uniform access to the 3--5~$\mu$m regime, where molecular opacity windows and thermal emission contribute substantially to both bolometric flux and atmospheric diagnostics. A recent SPHEREx based study has constructed the largest spectrophotometric sample of ultracool dwarfs to date, demonstrating the value of these data for population level studies of ultracool dwarf atmospheres and spectral diversity \citep{2026arXiv260422012G}.

The James Webb Space Telescope (JWST) has demonstrated the diagnostic power of 0.75--5~$\mu$m spectroscopy for cold brown dwarfs \citep{2024AJ....167....5L,2024AJ....167..237L,2025ApJ...980..230T,2025ApJSTu}, but necessarily in relatively small targeted samples. SPHEREx complements JWST by providing lower resolving power spectrophotometry for orders of magnitude more objects in a homogeneous survey, enabling population scale studies that were previously impractical.

In this work, we use SPHEREx Quick Release 2 (QR2) to construct the SPHEREx Ultracool Dwarf spectrophotometric Atlas (SUDA) and derive a uniform set of atmospheric parameters together with derived physical properties for a large field sample selected from UltracoolSheet \citep{UltracoolSheet2024}. The key advance is the combination of sample size and continuous 0.75--5~$\mu$m coverage, which allows us to examine atmospheric parameters, bolometric luminosities, an empirical spectrophotometric atlas, and carbon-bearing molecular behavior within one homogeneous sample. The following sections describe the data, sample construction, flux calibration, and fitting methodology, and then present the resulting atmospheric parameters, bolometric luminosities, evolutionary constraints, the empirical spectrophotometric atlas, and molecular index trends. Details of sample quality control and supporting analyses are provided in the appendices.

\section{Data and Sample} \label{sec:dataandsample}

This section describes the SPHEREx spectrophotometric data, the sample construction, the flux calibration, and the screening steps that define the final SUDA sample.

\subsection{SPHEREx Spectrophotometry} \label{subsec:spherexdata}

SPHEREx provides continuous 0.75--5.0~$\mu$m spectrophotometry suitable for studying ultracool dwarfs within one homogeneous survey. The SPHEREx spectrophotometric data used in this work were extracted from the NASA/IPAC Infrared Science Archive (IRSA) SPHEREx Quick Release Spectral Images using the IRSA SPHEREx Spectrophotometry Tool\footnote{\url{https://irsa.ipac.caltech.edu/applications/spherex/tool-spectrophotometry}} in November 2025. Observations are obtained with the Wide Field Linear Variable Filter (LVF) Imaging Spectrograph.

The Spectrophotometry Tool applies The Tractor forced-photometry framework to the spectral images at supplied target positions: it centers the position-dependent SPHEREx point-spread function (PSF) at the submitted target position and fits the flux amplitude, while the source centroid is not re-fitted or optimized from the SPHEREx image during the fit. Consequently, reliable propagated astrometry is important for high-proper-motion ultracool dwarfs. SPHEREx has an angular resolution of $6.15^{\prime\prime}$, corresponding to its photometric spatial resolution; this enables systematic studies of isolated sources while motivating careful treatment of unresolved multiples and crowded regions (Appendix~\ref{appendix:sample_qc_cal}).

IRSA began weekly releases of SPHEREx spectral-image data in July 2025 as Quick Release 1 (QR1). In October 2025, IRSA initiated Quick Release 2 (QR2), which reprocessed the full archive with improved calibration and data quality. All SPHEREx spectrophotometric measurements used in this work were extracted from QR2 images.

The SPHEREx spectrophotometry is sampled in 102 total channels distributed over six bands with continuous coverage from $0.75$ to $5.0~\mu$m. The four bands at $0.75$--$3.82~\mu$m provide low-resolution spectrophotometric sampling with typical resolving power $R \sim 40$, while the two bands at $3.82$--$5.0~\mu$m provide higher resolving power at $R \sim 110$--130. Preflight simulations predict $5\sigma$ point-source sensitivities of $m_{\rm AB}\sim18.5$--19 at $0.75$--$3.8~\mu$m and $m_{\rm AB}\sim16.6$--18 at $3.8$--$5.0~\mu$m \citep{2025ApJSCrill}. It is worth noting that each individual channel in the SPHEREx detectors has its own response function \citep{2026ApJS..284...10H}. Since the response functions were not released at the time of this work, we use approximate response-function treatments with the preliminary SPHEREx channel\footnote{we adopted the nominal wavelength limits from \url{https://irsa.ipac.caltech.edu/ibe/data/spherex/qr2/spectral_channels}, the IRSA QR2 spectral-channel calibration file. This preliminary SPHEREx channel is also available from Zenodo (DOI: \url{https://doi.org/10.5281/zenodo.20659256})} rather than channel-specific response models.

\subsection{Sample Selection from the UltracoolSheet} \label{subsec:sampleselection}

We selected the targets from the UltracoolSheet\footnote{\url{https://bit.ly/UltracoolSheet}}, which compiles known ultracool stars and brown dwarfs with astrometric, photometric, and ancillary physical information. For each source, we adopted the recommended J2000.0 coordinates and proper motions from the \texttt{ra\_j2000\_formula}, \texttt{dec\_j2000\_formula}, \texttt{pmra\_formula}, and \texttt{pmdec\_formula} columns. The J2000.0 coordinates were then linearly propagated to the October 2025 epoch using the adopted proper motions, with \texttt{pmra\_formula} treated as $\mu_{\alpha*}=\dot{\alpha}\cos\delta$. The propagated positions were submitted manually as point-source queries in the J2000.0 equatorial frame to the IRSA SPHEREx Spectrophotometry Tool, using the default optional settings: no extended-profile shape fit, a background-estimation-region value of \texttt{15}, and no explicit start- or end-time restriction. For repeatability, the full machine-readable catalog reports the adopted J2000.0 coordinates and proper motions together with the SPHEREx-epoch propagated right ascension and declination for each source (Table~\ref{tab:first30_trunc}). This procedure yielded SPHEREx spectrophotometry for 3164 ultracool objects.

Detailed procedures such as high S/N source screening and checks for blue-end contamination are provided in Appendix~\ref{appendix:sample_qc_cal}. After these screening steps, the final analysis sample contains 1675 ultracool objects.

\begin{figure*}[t]
  \centering
  \includegraphics[width=1\textwidth]{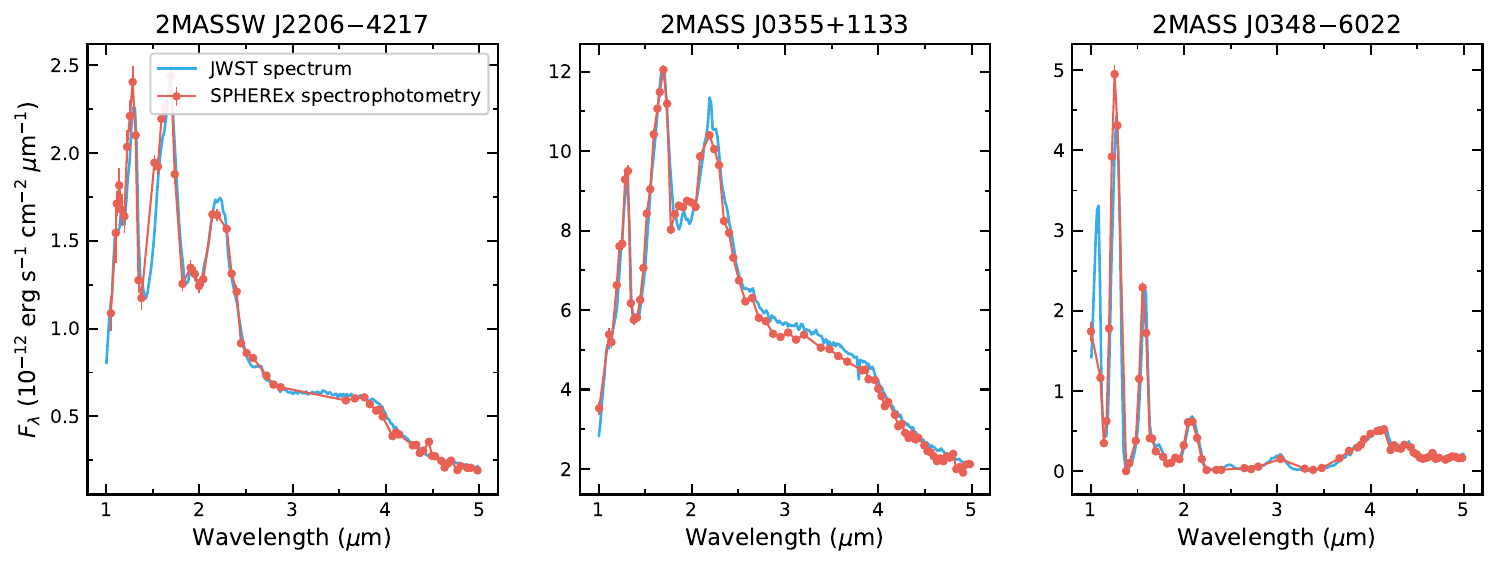}
  \caption{Comparison between the SPHEREx spectrophotometry and JWST spectra for 2MASSW~J2206$-$4217, 2MASS~J0355$+$1133, and 2MASS~J0348$-$6022. The SPHEREx measurements (red points), shown without additional flux rescaling, are compared with the corresponding JWST spectra (blue curves). The SPHEREx and JWST 1--5\,$\mu$m integrated fluxes differ by 1.0\%, 2.2\%, and 4.4\% for the three objects, respectively.}
  \label{fig:spherex_jwst_compare}
\end{figure*}

As an external check on the reliability of the retained SPHEREx spectrophotometry, we cross-matched our sample with objects analyzed in \citet{2025ApJSTu}. Figure~\ref{fig:spherex_jwst_compare} compares SPHEREx spectrophotometric measurements, without additional rescaling, against JWST spectra for three representative brown dwarfs. Because the full SPHEREx channel response functions are not available, we compute the integrated-flux comparison by averaging the JWST spectra over each SPHEREx channel using a uniform top-hat approximation defined by the channel wavelength limits. The SPHEREx and JWST 1--5\,$\mu$m integrated fluxes agree to within 5\% for all three objects (Figure~\ref{fig:spherex_jwst_compare}). In a channel-by-channel comparison, we define the normalized residual as $r_i=(f_{i,\rm SPHEREx}-f_{i,\rm JWST})/(\sigma_{i,\rm SPHEREx}^2+\sigma_{i,\rm JWST}^2)^{1/2}$. Combining the reported measurement uncertainties from both instruments gives mean absolute normalized residuals of 1.5, 4.2, and 2.3 for 2MASSW~J2206$-$4217, 2MASS~J0355$+$1133, and 2MASS~J0348$-$6022, respectively. The typical relative channel uncertainties are approximately 4.5\% for SPHEREx and 0.18\% for JWST. These formal residuals do not account for cross-instrument absolute flux-calibration systematics, uncertainties associated with the approximate SPHEREx channel-response treatment, or possible source variability between the observing epochs.

\begin{figure*}[t]
  \centering
  \includegraphics[width=0.88\textwidth]{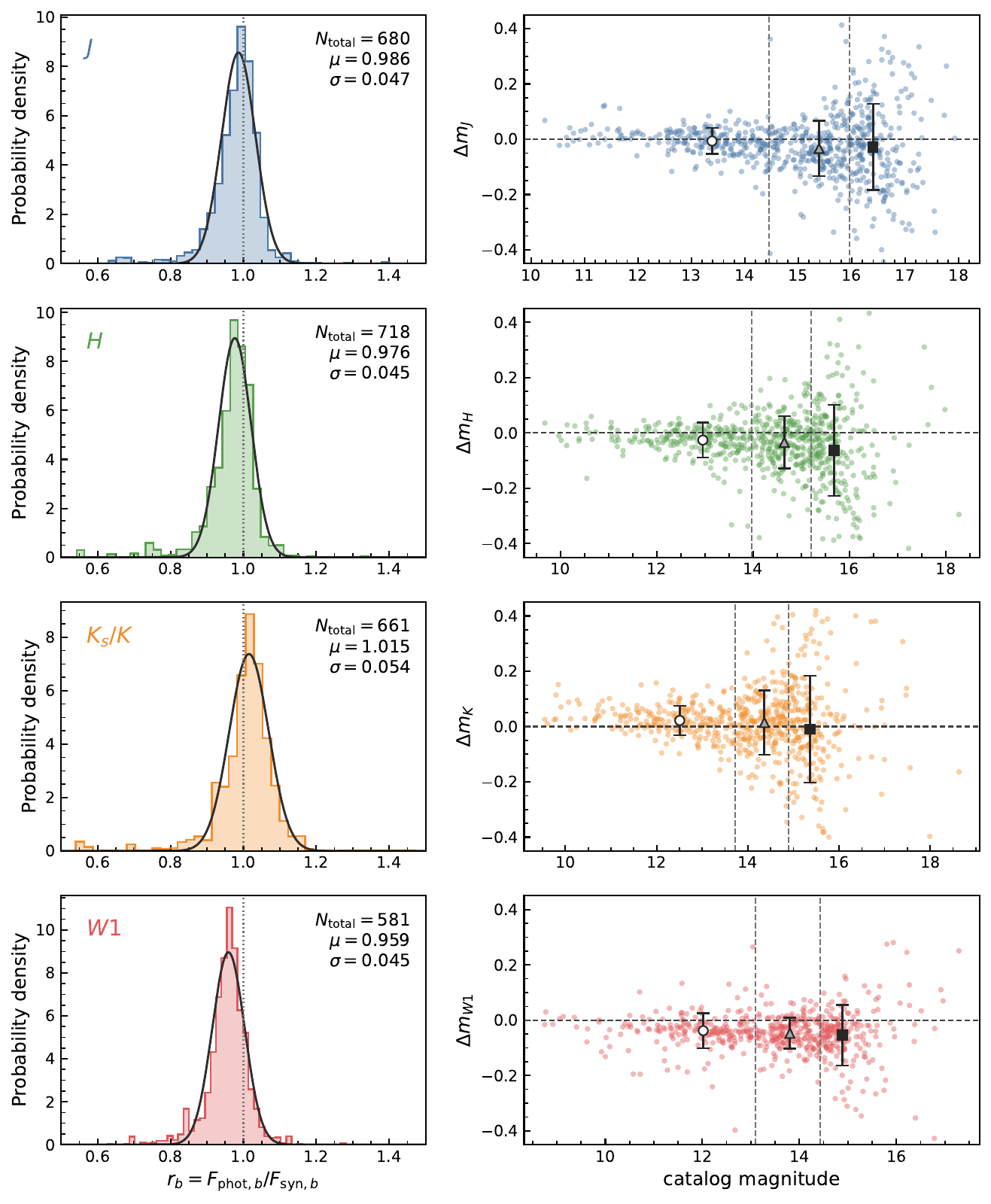}
  \caption{
  Individual-band consistency diagnostics for the retained $J$, $H$,
  $K_s/K$, and $W1$ measurements. The left column shows the distributions
  of the band-level flux ratio
  $r_b=F_{\mathrm{phot},b}/F_{\mathrm{syn},b}$ after requiring greater
  than 90\% response-weighted SPHEREx coverage. The right
  column shows the magnitude residual,
  $\Delta m=m_{\rm SPHEREx}-m_{\rm catalog}$, as a function of catalog
  magnitude. The vertical dashed lines divide each band into equal-number
  bright, middle, and faint tertiles, and the overplotted symbols and error
  bars show the mean and standard deviation of $\Delta m$ in each tertile.
  The scatter is approximately 0.05~mag in the bright tertiles of all four
  bands. In the middle tertiles, it increases to approximately 0.1~mag in
  $J$, $H$, and $K_s/K$, while remaining near 0.05~mag in $W1$. In the
  faint tertiles, the scatter reaches approximately 0.15--0.20~mag in the
  near-infrared bands and approximately 0.1~mag in $W1$.
  }
  \label{fig:R_dis_band}
\end{figure*}

\subsection{Flux Calibration}
\label{subsec:flux_cali}

We use external photometry to evaluate the absolute flux consistency of the SPHEREx spectrophotometry and to correct sources with significant gray flux-scale offsets. For each source, we use 2MASS $J$, $H$, and $K_s$ when available, MKO $J$, $H$, and $K$ otherwise, and WISE $W1$ to anchor the longer-wavelength flux scale. We do not compute synthetic WISE $W2$ magnitudes because the $W2$ response extends beyond the 5.0~$\mu$m wavelength limit of the SPHEREx spectrophotometry, leaving part of the filter bandpass uncovered.

For each external photometric band $b$, we calculate a synthetic SPHEREx flux by integrating the retained SPHEREx channels over the corresponding filter response curve, without interpolating across missing channels. Each SPHEREx channel is treated as a constant $F_\nu$ measurement over its official wavelength interval. The synthetic flux is given by
\begin{equation}
\label{eq:synth-fnu}
F_{\mathrm{syn},b} = \frac{\sum_j F_{\nu,j} A_{j,b}}{\sum_j A_{j,b}},
\end{equation}
where $A_{j,b} = \int_{\Delta\lambda_j} T_b(\lambda)\lambda^{-1}\,d\lambda.$ Here $F_{\nu,j}$ is the SPHEREx flux density in channel $j$, $\Delta\lambda_j$ is the wavelength interval of that channel, and $T_b(\lambda)$ is the transmission curve of external band $b$. The synthetic-flux uncertainties are propagated in quadrature using the same normalized weights, $A_{j,b}/\sum_j A_{j,b}$. We retain only bands with response-weighted SPHEREx coverage $\sum_j A_{j,b}/A_{b,\rm full}$ greater than 90\%, where $A_{b,\rm full}$ is the corresponding full-filter integral.

For each retained band, we define the photometric flux ratio as $r_b=F_{\mathrm{phot},b}/F_{\mathrm{syn},b}$. The individual band ratios are then combined into a single source-level multiplicative scale factor,
\begin{equation}
\label{eq:fcal}
f_{\rm cal} = \frac{\sum_b w_b r_b}{\sum_b w_b},
\end{equation}
where the weights $w_b$ account for the catalog-photometry uncertainty, the propagated synthetic-flux uncertainty, and the empirical scatter of the ratio distribution in each band. We add a 0.01~mag uncertainty floor in quadrature to the catalog magnitude uncertainty and include a 1\% relative uncertainty floor in the weighted combination.

Figure~\ref{fig:R_dis_band} shows the individual-band consistency diagnostics. The $J$, $H$, and $K_s/K$ ratios are centered close to unity, whereas the $W1$ distribution has an empirical center of $\mu_{W1}=0.959$. The residual scatter increases toward fainter magnitudes. The overall dispersion is approximately 5\% in flux, broadly consistent with the uncertainty reported in the SPHEREx Explanatory Supplement\footnote{\url{https://irsa.ipac.caltech.edu/data/SPHEREx/docs/SPHEREx_Expsupp_QR.pdf}}. Because our analysis is restricted to a quality-controlled subsample, this value should not be interpreted as representative of the full SPHEREx source population, for which the uncertainties may be larger \citep{2026arXiv260422012G}. Several effects may contribute to the observed mean $W1$ offset. The approximate filter response functions adopted in this work may not fully represent the true instrumental response, introducing biases in the synthetic flux calculation. Systematic uncertainties in background subtraction, either from the SPHEREx extraction or catalog photometry, may also affect the flux ratio. In addition, incomplete SPHEREx channel coverage within the $W1$ bandpass can introduce an uncertainty of approximately 1\% based on a linear interpolation test for missing channels. Finally, the spectrophotometry used here was extracted before the March 31, 2026 correction to the QR2 PSF headers and the SPHEREx Spectrophotometry Tool\footnote{\url{https://irsa.ipac.caltech.edu/data/SPHEREx/docs/psfhdrerr.html}}. Tests with updated QR2 spectrophotometry for a subset of sources show that $r_{W1}$ shifts toward unity by approximately 1\%, suggesting that calibration updates may account for part of the original offset.

Before deciding whether an individual source requires flux calibration, we remove the band-dependent offsets measured from the full retained sample. For each band, the ratio used in the step is therefore $r_b/\mu_b$, where $\mu_b$ is the adopted center of the ratio distribution for that band. We then evaluate the same weighted average as in Equation~(\ref{eq:fcal}). For sources with at least three retained bands, we require all band ratios to lie on the same side of unity and their weighted average to differ from unity by more than 10\%. For sources with two retained bands, we adopt a 15\% threshold. Sources with only one accepted band are not considered.

These criteria identify 47 sources requiring flux calibration, including 26 sources with at least three retained bands and 21 sources with two retained bands. For these sources, the SPHEREx spectrophotometry is multiplied by the corresponding value of $f_{\rm cal}$. The synthetic photometry listed in Table~\ref{tab:first30_trunc} is calculated from the uncorrected SPHEREx fluxes, and sources subsequently corrected by this procedure are marked with an asterisk.

\section{Atmospheric Model Fitting}\label{sec:model_fitting}

We use a uniform two-stage fitting pipeline to derive atmospheric parameters for the full SPHEREx sample. We first perform minimum $\chi^2$ fitting on model grids to identify an initial solution for each source, then refine that solution with nested sampling Bayesian inference. 

\subsection{Initial Grid-Based $\chi^2$ Fitting}\label{subsec:initial_grid_fit}

We first use grid-based $\chi^2$ fitting to obtain an initial atmospheric solution for each source. Our primary model grid is the Spectral ANalog of Dwarfs (SAND; \citealt{2024RNAASAlvarado}), which spans 700--4000~K and covers the broad surface gravity and metallicity range required by this sample. Its demonstrated performance across L/T transition spectra \citep{2025ApJSTu} also makes it well suited to our heterogeneous data set.

We evaluated $\chi^2$ for all recommended\footnote{For the definition of ``recommended'', please refer to \url{https://zenodo.org/records/11582126}.} SAND spectra after convolving models to the SPHEREx resolution ($R=40$ over $0.7$--$3.8~\mu$m and $R=110$ over $3.8$--$5.0~\mu$m). Following Equations~(6) and~(7) in \citet{2023ApJLBeiler}, we solve for a wavelength independent flux scaling and adopt the minimum $\chi^2$ parameter set as the preliminary solution for each object.

For a subset of very cool sources, we also fit with the ATMO2020++ model grid (\citealt{2023AJMeisner}), which is optimized for late T/Y regimes and now spans $[\mathrm{M/H}] = -1.0$ to $+0.3$. For objects with preliminary SAND fits of $T_{\mathrm{eff}} \leq 1300$~K, we repeat the same minimum $\chi^2$ procedure with ATMO2020++ and adopt the model family with the smaller reduced $\chi^2$.

\subsection{Bayesian Parameter Inference with Nested Sampling}\label{subsec:nested_sampling}

We then refine those initial solutions with Bayesian inference to obtain the final parameter estimates and uncertainties. After the initial $\chi^2$ step, we use the \texttt{UltraNest} nested sampling framework \citep{untranest2016, ultranest2019,ultranest} for each source. For each object, we use the atmospheric model family (SAND or ATMO2020++) preferred by the initial optimization.

For a given model grid, we linearly interpolate in all available parameter dimensions in logarithmic flux space. For SAND, the native grid at $T_{\mathrm{eff}} < 2500$~K covers $\log g = 4.0$--6.0. For sources with initial solutions at $T_{\mathrm{eff}} < 2500$~K and $\log g = 4.0$, we extend the grid to $\log g = 3.5$ by linear extrapolation in logarithmic flux space using the two lowest $\log g$ points at fixed $T_{\mathrm{eff}}$, [M/H], and [$\alpha$/Fe]. Extrapolations below $\log g = 3.5$ are not allowed.

Posterior sampling is performed under a Gaussian likelihood. We define an effective variance at each wavelength point $i$ as
\begin{equation}
  \sigma_{\mathrm{eff},i}^2 = \sigma_i^2 + \left( f\,F_{\mathrm{obs},i} \right)^2 ,
\end{equation}
where $\sigma_i$ is the reported observational uncertainty and $f$ is a fractional error inflation parameter. The log likelihood is then given by
\begin{equation}
  \ln \mathcal{L} =
  -\frac{1}{2} \sum_{i}
  \left[
    \frac{\left(F_{\mathrm{obs},i} - F_{\mathrm{mod},i}\right)^2}
    {\sigma_{\mathrm{eff},i}^2}
    + \ln \left( 2\pi \sigma_{\mathrm{eff},i}^2 \right)
  \right].
\end{equation}
Here $F_{\mathrm{obs},i}$ is the observed flux at wavelength point $i$, and $F_{\mathrm{mod},i}$ is the corresponding model flux after convolution to instrumental resolution and flux scaling.

Prior ranges are informed by the initial $\chi^2$ solution: $T_{\mathrm{eff}} \pm 200$~K, $\log g \pm 1$~dex, and uniform priors over the full model grid ranges for remaining parameters (including [M/H] and, when available, [$\alpha$/Fe]). For the flux scaling parameter $\log(R^2/D^2)$, we use a uniform prior centered on the initial $\chi^2$ value with half-width 0.1 dex. The fractional error inflation parameter is assigned a uniform prior of $0 \leq f \leq 0.2$.

From the nested sampling posteriors, we adopt medians as parameter estimates and the 16th/84th percentiles as uncertainties.

\section{Model Fitting Results}\label{sec:model_results}

This section presents the atmospheric parameters, bolometric luminosities, spectral types, and evolutionary quantities derived from the SPHEREx spectrophotometry and the best-fitting SAND/ATMO2020++ atmospheric models. A sample of the catalog is provided in Appendix~B (Table~\ref{tab:first30_trunc}), including the best fit atmospheric parameters, adopted model family, distance, literature spectral type, and spectral type from this work. The spectral types from this work are obtained with SPLAT \citep{2017ASInC..14....7B} by minimizing $\chi^2$ against the standard spectral templates, and are left blank when the SPHEREx spectrophotometry has insufficient S/N or incomplete blue wavelength coverage for a reliable match. The classifications agree at approximately the one subtype level, with a mean offset of $-0.4$ subtype, and a median absolute deviation (MAD) of $1.0$ subtype.

\subsection{Bolometric Fluxes and Bolometric Luminosities}
\label{subsec:fbol_lbol}

\subsubsection{Bolometric Flux Estimation}
\label{subsec:fbol}

We derive bolometric fluxes by integrating composite SEDs constructed from the SPHEREx spectrophotometry and the best-fitting atmospheric models. Within the SPHEREx wavelength coverage, we use the observed spectrophotometric SED directly; outside this range, the best-fitting model provides the blue and red extrapolations. Because the full pixel-level response functions are not yet publicly available, we do not forward-model the composite SED through the SPHEREx channel responses. We therefore approximate the SPHEREx measurements as spectral flux points and integrate the resulting composite SED to obtain $F_{\mathrm{bol}}$. This treatment should provide a useful proxy for $F_{\mathrm{bol}}$, but the slight overlap among neighboring SPHEREx photometric bandpasses may leave small spectral-type-dependent systematics that would be better captured by a full response-function forward model.

\begin{figure}[t]
  \centering
  \includegraphics[width=\columnwidth]{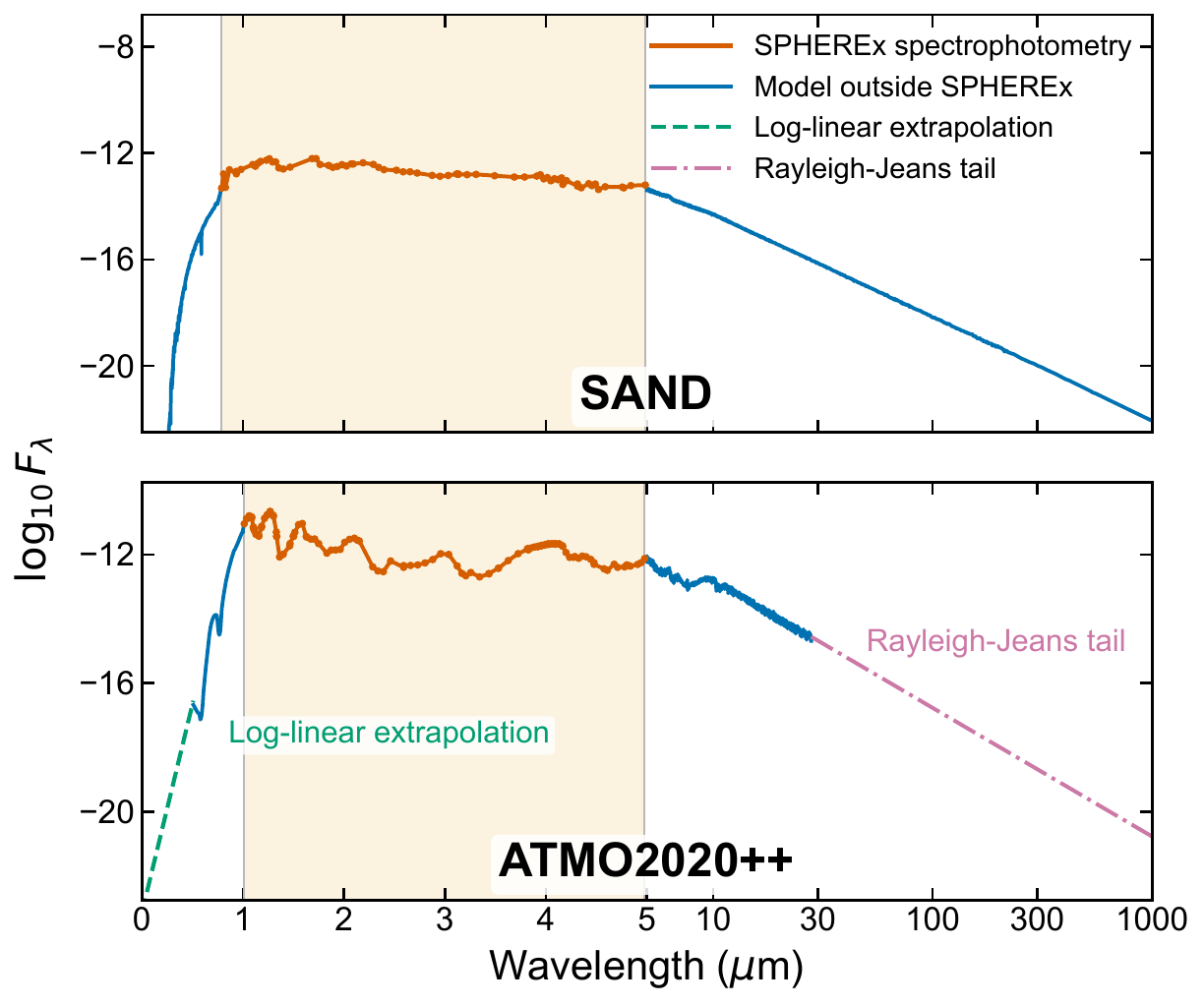}
  \caption{Schematic examples of the composite SEDs used to estimate $F_{\mathrm{bol}}$ for one SAND-fitted source and one ATMO2020++-fitted source. The two panels share the same wavelength axis, which is linear from 0--5~$\mu$m and logarithmically compressed from 5--1000~$\mu$m, but use independent $\log_{10} F_\lambda$ axes. The SPHEREx 102-channel measurements define the observed contribution over the measured wavelength range and replace the model there. Outside the SPHEREx range, the scaled best fitting atmospheric model supplies the remaining flux contribution. For the ATMO2020++ example, the short-wavelength log-flux linear extrapolation and long-wavelength Rayleigh--Jeans tail are shown separately.}
  \label{fig:composite_fbol_sed}
\end{figure}

Figure~\ref{fig:composite_fbol_sed} illustrates how the observed SPHEREx spectrophotometry and the model-only wavelength segments are combined to form the composite SEDs used for $F_{\mathrm{bol}}$ estimation. For sources for which the SAND model provides the best fit, the observed SPHEREx measurements replace the model over the observed range in the composite SED. Because the SAND grid spans $\sim 0.1$--$999\,\mu$m, no extra end extrapolation is required.

For sources for which the ATMO2020++ model provides the best fit, we use the same integration scheme but account for the more limited model coverage: a log-flux linear extrapolation is applied blueward of the model grid, and a Rayleigh--Jeans tail is appended redward of the model grid.

Uncertainties in $F_{\mathrm{bol}}$ are propagated with Monte Carlo sampling. In each realization, atmospheric parameters and flux scaling are drawn from their posteriors, observed fluxes are perturbed within reported errors, and an additional 5\% absolute calibration term is applied (Section~\ref{subsec:flux_cali}). The perturbed composite SED is then integrated in the same way as the fiducial SED, and the resulting distribution of $F_{\mathrm{bol}}$ values defines the uncertainty.

\subsubsection{Bolometric Luminosities from Parallax}\label{subsec:lbol_parallax}

For sources with parallaxes, we convert the measured bolometric fluxes into direct estimates of $L_{\mathrm{bol}}$. Bolometric luminosities are computed from the bolometric flux as
\begin{equation}
  L_{\mathrm{bol}} = 4\pi d^2 F_{\mathrm{bol}},
\end{equation}
where $d$ is the distance to the source. Distance uncertainties are propagated by sampling $d$ for each Monte Carlo realization in the conversion from $F_{\mathrm{bol}}$ to $L_{\mathrm{bol}}$. Bolometric luminosities are reported in units of $\log(L_{\mathrm{bol}}/L_\odot)$, adopting a solar bolometric luminosity of $L_\odot = 3.828\times10^{33}\,\mathrm{erg\,s^{-1}}$ \citep{2015arXiv151006262M}.

\subsubsection{Machine Learning Luminosity Estimates}
\label{subsec:ml_lbol}

For sources without parallaxes, we estimate $L_{\mathrm{bol}}$ from atmospheric parameters with XGBoost. Because $F_{\mathrm{bol}}$ cannot be converted to $L_{\mathrm{bol}}$ without a distance estimate, we adopt a supervised machine learning estimator that predicts $\log(L_{\mathrm{bol}}/L_\odot)$ from atmospheric parameters instead of relying on empirical relations between spectral type and absolute magnitude \citep{2012ApJSDupuy,2023ApJSanghi}.

We use an XGBoost regressor to map $(T_{\mathrm{eff}},\,\log g,\,[\mathrm{M/H}])$ to $\log(L_{\mathrm{bol}}/L_\odot)$. Because these inputs are available for the full sample, the estimator can be applied uniformly. This ML estimate is conceptually distinct from the luminosity calculation based on parallax. XGBoost is a gradient boosted decision tree algorithm that has been widely adopted in astronomy for regression and classification problems owing to its flexibility, robustness to nonlinear parameter dependencies, and strong performance on heterogeneous data sets \citep{2023ApJS..267....8A,2024ApJ...974..138Z,2025ApJS..280...13W,2025AJ....170..360Z}.

For each source without parallax, we report the luminosity predicted by the ML model, $\log(L_{\mathrm{bol}}/L_\odot)$, and its uncertainty from Monte Carlo propagation of atmospheric parameter uncertainties through the trained model. These ML luminosities are used only for sources without parallax. Details of model training and validation are given in Appendix~\ref{appendix:ml_lbol_details}. The trained XGBoost estimator is publicly available from Zenodo (DOI: \url{https://doi.org/10.5281/zenodo.20659256}).

\subsubsection{Bolometric Luminosity Comparison}\label{subsec:lbol_comparison}

\begin{figure}[h]
  \centering
  \includegraphics[width=0.48\textwidth]{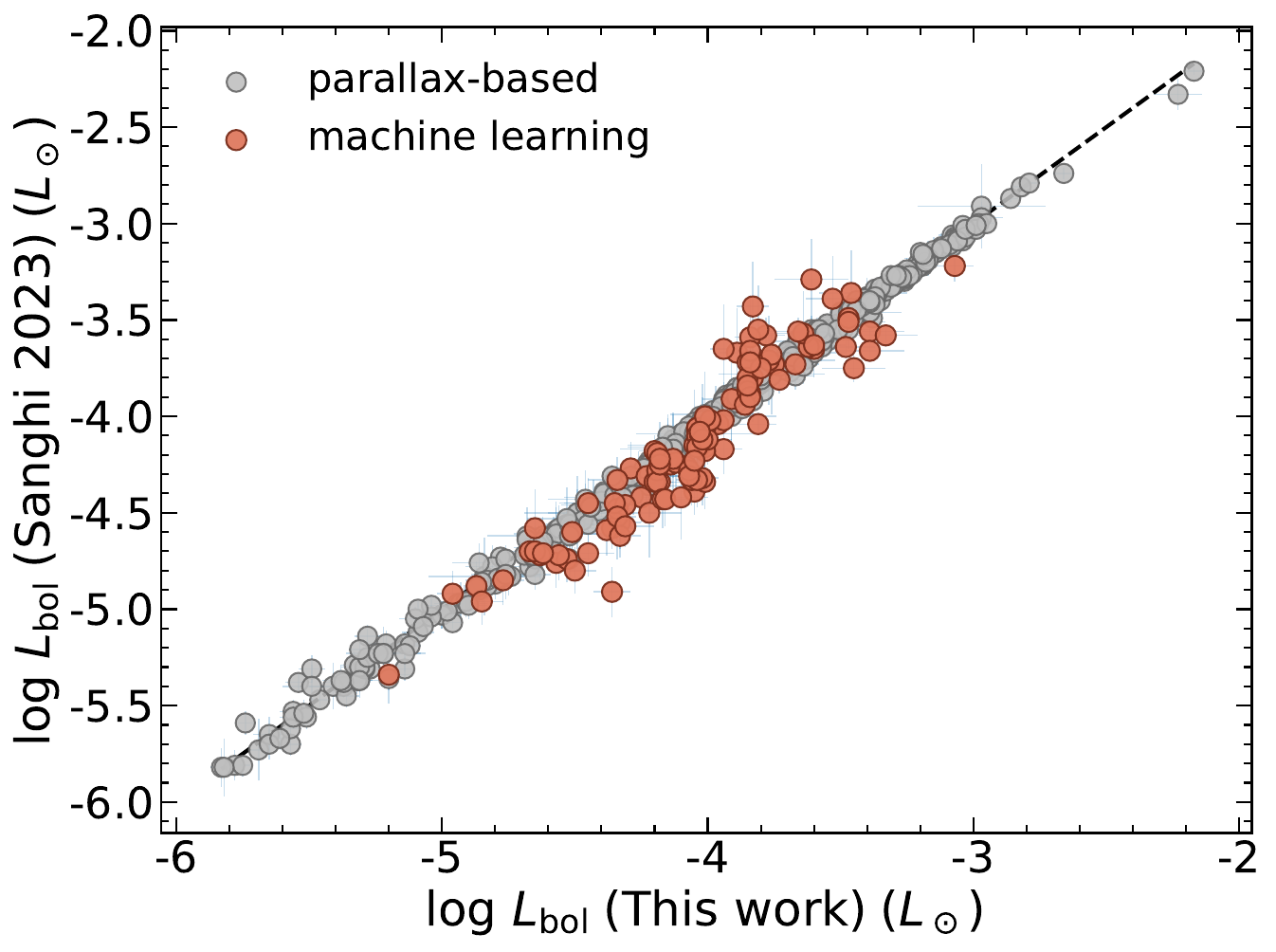}
  \caption{Comparison of $\log L_{\mathrm{bol}}$ between this work and \citet{2023ApJSanghi} for 592 matched sources. The sample includes 117 sources without parallax, for which this work uses estimates from the ML model and \citet{2023ApJSanghi} uses relations between absolute magnitude and $L_{\mathrm{bol}}$.}
  \label{fig:lbol_crossmatch}
\end{figure}

We compare our luminosities with literature values to assess the consistency of both the estimates based on parallax and those from XGBoost. Specifically, we compare our $\log L_{\mathrm{bol}}$ measurements with values from \citet{2023ApJSanghi} for 592 matched sources. Of these, 117 lack parallax measurements; our values for that subset come from the ML estimator, while \citet{2023ApJSanghi} used empirical relations for absolute magnitude. Figure~\ref{fig:lbol_crossmatch} shows broad consistency between the two catalogs, with larger scatter in the subset without parallax as expected from the different inference strategies. The 475 parallax-based sources have a mean difference of $0.016$~dex and a MAD of $0.035$~dex. The 117 ML-based sources have a larger mean difference of $0.091$~dex and a MAD of $0.090$~dex, reflecting the additional scatter introduced when luminosities must be inferred without parallaxes.

\subsection{Parameters Derived from Evolutionary Models}
\label{sec:evolutionary_parameters}

\begin{figure*}[t]
  \centering
  \includegraphics[width=0.95\textwidth]{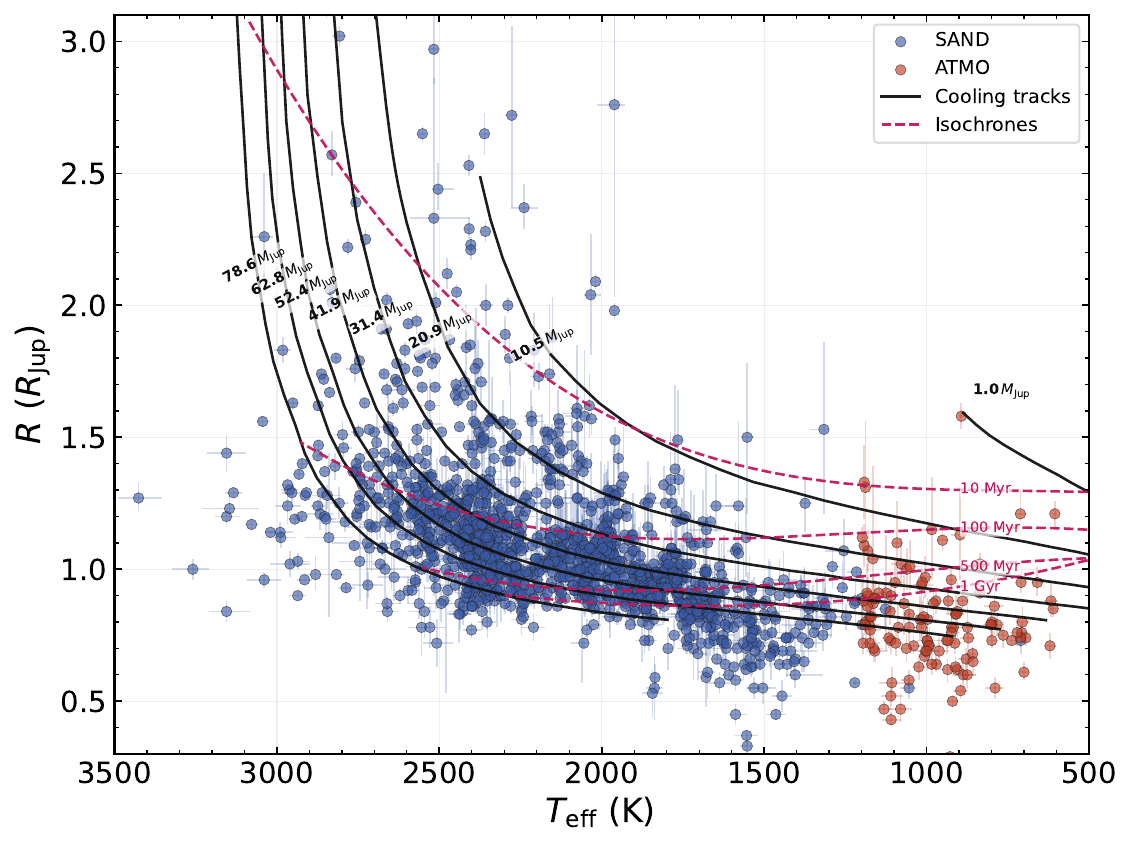}
  \caption{Radius as a function of effective temperature for SUDA sources, overlaid with representative C23 evolutionary tracks. Solid C23 curves show cooling tracks, with masses labeled in the panel; dashed colored curves show selected isochrones, with ages labeled in the panel. Blue and orange points indicate sources best fitted by SAND and ATMO2020++, respectively.}
  \label{fig:radius_teff_c23}
\end{figure*}

We adopt the $(T_{\mathrm{eff}},R)$ combination as the input to the C23 substellar evolutionary tracks \citep{2023A&AChabrier} to estimate mass and age. For the subset of sources with parallaxes, we also use the same tracks to estimate evolutionary $\log g$. Tests using $(\log L_{\mathrm{bol}},R)$ give broadly consistent results, but $(T_{\mathrm{eff}},R)$ has weaker covariance between the two input quantities and therefore suffers less degeneracy in the evolutionary grid.

For sources with parallaxes, the radius used for this interpolation is obtained from the fitted spectral scaling, $\log(R^2/D^2)$, together with the parallax-based distance. For sources without parallaxes, the radius can only be inferred indirectly from model-based quantities associated with the ML estimate of $L_{\mathrm{bol}}$. Thus, we estimate evolutionary $\log g$ only for sources with parallaxes.

Figure~\ref{fig:radius_teff_c23} shows the resulting source distribution in the $(T_{\mathrm{eff}},R)$ diagram together with representative C23 evolutionary tracks. Most objects lie within, or close to, the region covered by the model grid. For sources that fall slightly outside the convex hull of the C23 grid, we allow a local affine extrapolation based on the nearest C23 grid points in the normalized $(T_{\mathrm{eff}},R)$ parameter space. In this diagram, the evolutionary tracks provide the mapping between the measured radius, the adopted effective temperature, and the inferred physical parameters. At fixed mass, the tracks evolve from larger radii at young ages toward smaller radii at older ages as the objects contract and cool. The sample follows the broad trend expected from these C23 tracks.

Uncertainties are propagated from both input quantities. The radius uncertainty includes the contributions from the distance and the fitted $\log(R^2/D^2)$ scaling. For $T_{\mathrm{eff}}$, we combine the formal fitting uncertainty with an additional conservative $120$~K uncertainty term. For each source, we sample the input distributions and report the median value and the 16th and 84th percentiles of the resulting C23 mass, age, and evolutionary $\log g$ (for those with parallaxes) distributions in Table~\ref{tab:first30_trunc}. For the final C23 estimates, the median fractional mass uncertainty for the full set is $21\%$, and the median age uncertainty is $0.25$~dex. For the subset of sources with parallaxes, the corresponding median uncertainties are $17\%$ in mass and $0.15$~dex in age.

\begin{figure*}[t]
  \centering
  \includegraphics[width=0.95\textwidth]{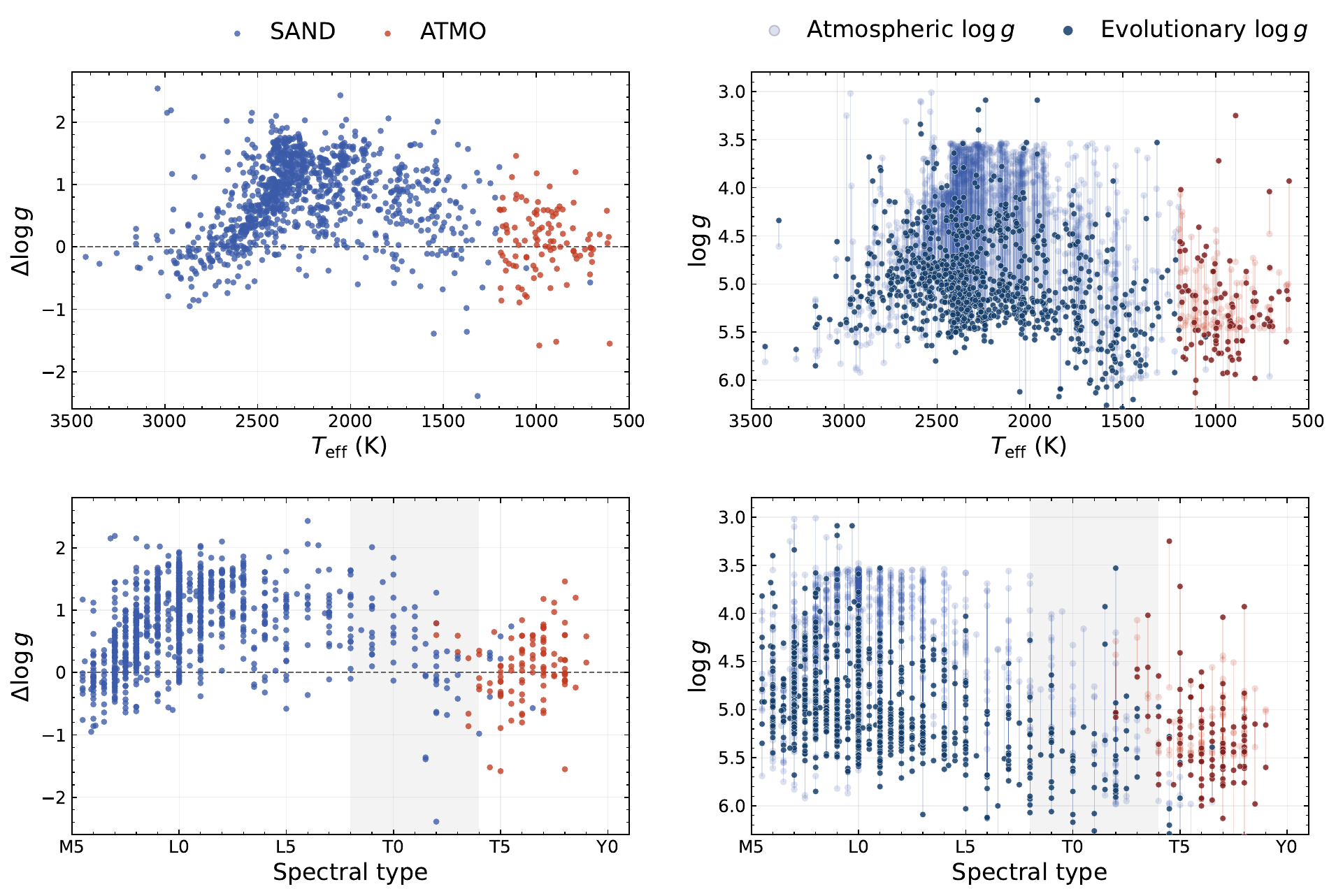}
  \caption{Left: difference between the C23 evolutionary $\log g$ and the atmospheric $\log g$. Right: atmospheric and evolutionary $\log g$, with atmospheric values shown as transparent background points and evolutionary values shown with darker points. The upper panels use $T_{\mathrm{eff}}$ as the x axis, and the lower panels repeat the comparison as a function of spectral type. In the lower panels, the shaded region marks the approximate L/T transition.}
  \label{fig:logg_atm_evol_comparison}
\end{figure*}

\subsubsection{Comparison of Atmospheric and Evolutionary Surface Gravities}
\label{subsec:gravity_comparison}

The evolutionary $\log g$ inferred from the C23 tracks is not identical to the atmospheric $\log g$ from spectral fitting. Figure~\ref{fig:logg_atm_evol_comparison} compares these two gravity estimates directly. The left column shows $\Delta\log g \equiv \log g_{\rm evol}-\log g_{\rm atm}$, while the right column shows the two estimates in the $\log g$ diagram. The upper row uses $T_{\mathrm{eff}}$ as the x axis, and the lower row repeats the same comparison against spectral type.

The most prominent difference occurs at $T_{\mathrm{eff}}\simeq1700$--2500~K. In this temperature range, the atmospheric gravities from spectral fitting are systematically lower than the evolutionary gravities, with a median offset of $\log g_{\rm evol}-\log g_{\rm atm}\simeq1.05$~dex. The spectral-type projection shows that the large positive offsets are concentrated across the L-dwarf sequence and remain relevant near the L/T transition, where condensate clouds strongly shape the near-infrared spectral morphology. Outside this interval, the offsets are smaller and do not show the same systematic sign. This pattern suggests that the low atmospheric gravities inferred for many sources in the cloudy L-dwarf regime should be interpreted with caution.

The likely explanation is the model dependence in the cloudy L-dwarf regime. Cloud opacity and cloud clearing across the L sequence and L/T transition \citep[e.g.,][]{2012ApJ...756..172M} can alter the near-infrared morphology used in the atmospheric fits, and non-equilibrium CO/CH$_4$ chemistry \citep[e.g.,][]{2014ApJ...797...41Z} can add further model dependence. When these cloud and chemistry effects are not fully captured by the atmospheric grids, the fitting procedure can partly compensate through $\log g$ and [M/H], especially because the SAND models span a broad metallicity range. Thus, the atmospheric--evolutionary gravity mismatch is likely driven by coupled degeneracies among cloud properties, $\log g$, metallicity, and possible non-equilibrium chemistry associated with vertical mixing in L-dwarf atmospheres.

The detailed spectral-type dependence of this mismatch is nevertheless model dependent. For example, the Sonora Diamondback comparison in Figure 11 of \citet{2026AJ....171..198M} does not show an equally prominent gravity discrepancy at the L/T transition; in that analysis, the larger atmospheric--evolutionary $\log g$ offsets occur mainly near the warmer early-L sequence around L0 and among cooler objects later than about T2. Thus, the concentration of large positive offsets in our comparison should be interpreted as a SAND-dependent behavior of the atmospheric fits.

\subsection{The Spectrophotometric Atlas}\label{sec:spectral_atlas}

\begin{figure*}[t]
  \centering
  \includegraphics[width=0.95\textwidth,height=0.95\textheight]{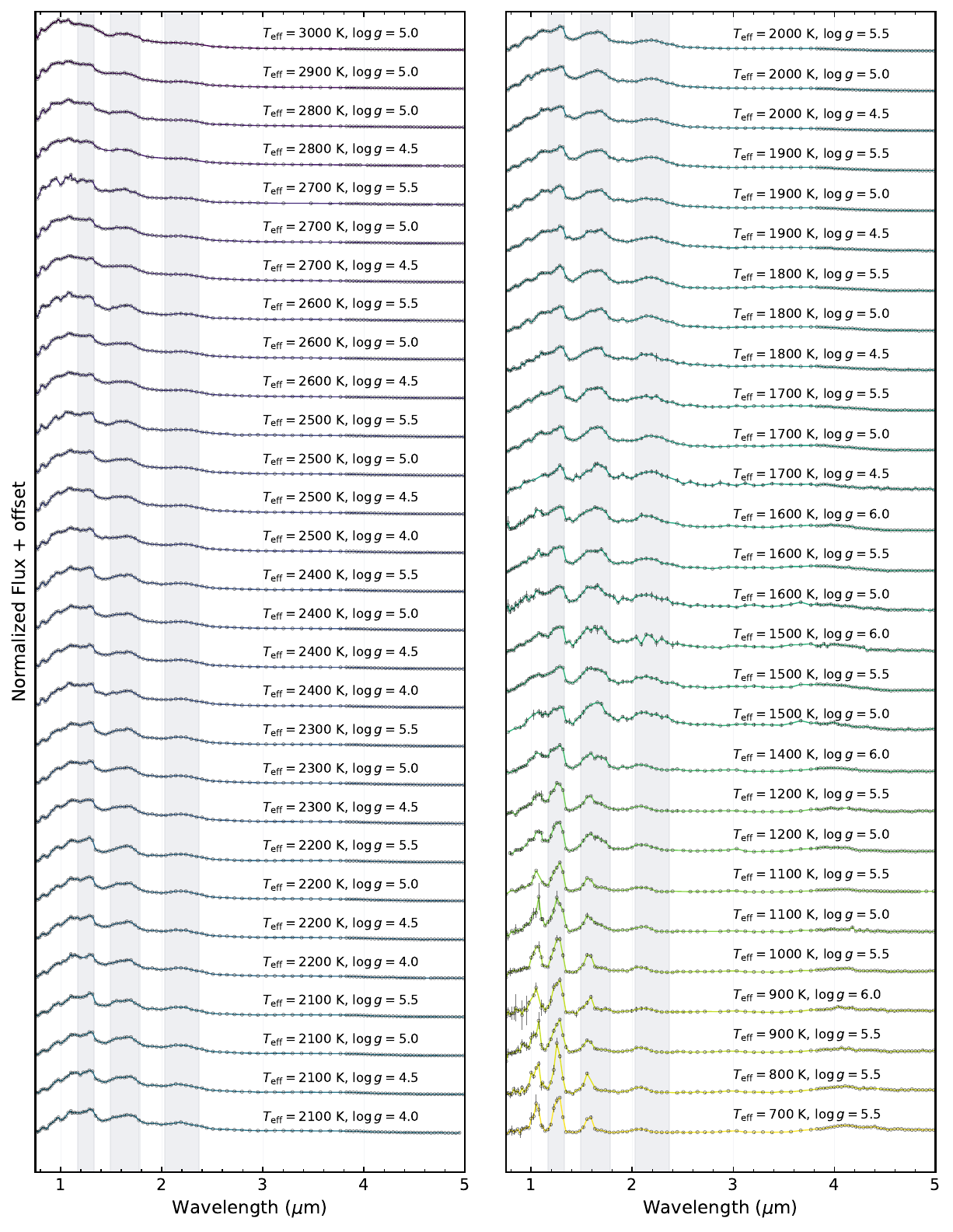}
  \caption{Empirical SPHEREx Ultracool Dwarf spectrophotometric Atlas constructed from the $T_{\mathrm{eff}}$ and $\log g$ bins. Each point is the median of the normalized source fluxes at one released SPHEREx channel in a given parameter bin, with vertical error bars. Labels give the adopted parameters of each bin as $(T_{\mathrm{eff}}, \log g)$, where $T_{\mathrm{eff}}$ is from atmospheric model fitting and $\log g$ is from the C23 evolutionary model. Gray shaded regions mark the $J$, $H$, and $K$ bands. This spectrophotometric atlas is available from Zenodo (DOI: \url{https://doi.org/10.5281/zenodo.20659256}).}
  \label{fig:spectral_atlas}
\end{figure*}

We present an empirical SPHEREx atlas built directly from the retained 102-channel SPHEREx spectrophotometry. Sources are grouped using the $T_{\mathrm{eff}}$ fitted with atmospheric models and the $\log g$ inferred from evolutionary models, with bin widths of $\Delta T_{\mathrm{eff}}=100$~K and $\Delta \log g=0.5$ centered on the atlas grid points. Each source is normalized by the median flux of the 1.284 and 1.315~$\mu$m channels. Each atlas point is then defined as the median of the retained normalized source measurements. Bins with fewer than six input sources are excluded, leaving 57 atlas bins constructed from 986 input sources. Individual atlas channels are retained when at least three normalized source measurements contribute, so a small number of bins do not contain all 102 channels.

This construction is intended to preserve the characteristic morphology of the observed SPHEREx spectrophotometry associated with different regions of the adopted $T_{\mathrm{eff}}$ and $\log g$ parameter space, while suppressing random noise, residual artifacts, and source specific peculiarities. Because the atlas is assembled from normalized 102-channel observations by taking the median in each bin, it should be interpreted primarily as an empirical reference sequence in the observed 0.75--5~$\mu$m.

Figure~\ref{fig:spectral_atlas} presents the resulting 102-channel atlas as two vertically offset spectrophotometric sequences, ordered by decreasing $T_{\mathrm{eff}}$. With this ordering, the dominant temperature driven evolution of the SPHEREx spectrophotometric morphology is seen directly: the broad molecular bands strengthen and change shape systematically from the late M regime through the T/Y regime. The labels retain the adopted $\log g$ value for each bin, while the ordering emphasizes the temperature driven morphology of the atlas.

\section{Molecular-Index Trends and Model Interpretation}
\label{sec:molecular_indices}

\subsection{Empirical Molecular-Index Sequences}
\label{subsec:empirical_molecular_indices}

We use molecular indices to characterize the temperature-dependent spectrophotometric sequence of the SUDA sample and to identify secondary atmospheric effects. H$_2$O and CH$_4$ primarily trace temperature-dependent molecular chemistry \citep{Geballe2002Tclass}, whereas CO and CO$_2$ can additionally respond to vertical mixing, metallicity, and non-equilibrium chemistry in the 3--5~$\mu$m region \citep{2007ApJ...655.1079L,2009ApJ...695..844G}.

\begin{figure*}[t]
  \centering
  \includegraphics[width=0.95\textwidth]{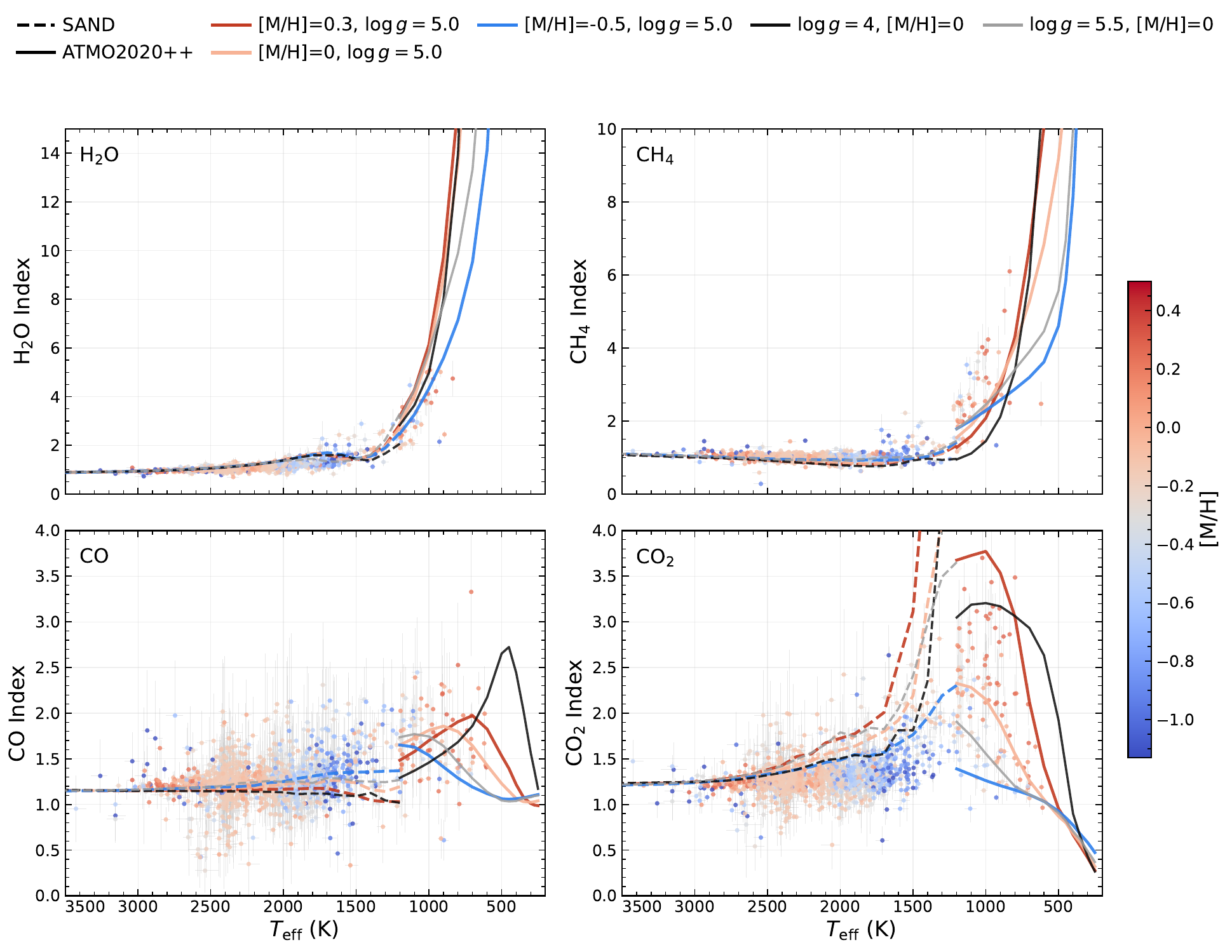}
  \caption{Molecular-index trends as a function of effective temperature for H$_2$O, CH$_4$, CO, and CO$_2$. The observed sources are colored by their fitted $[\mathrm{M/H}]$. The dashed and solid curves show the SAND and ATMO2020++ model sequences, respectively. At fixed $\log g=5.0$, the metallicity sequences are shown in red for $[\mathrm{M/H}]=+0.3$, medium black for $[\mathrm{M/H}]=0.0$, and blue for $[\mathrm{M/H}]=-0.5$. At fixed solar metallicity, the additional gravity sequences are shown in dark black for $\log g=4.0$ and light gray for $\log g=5.5$.}
  \label{fig:mol_index_teff}
\end{figure*}

We compute all four indices using the continuum-to-feature ratio $I_{\rm mol}=\langle F_{\lambda,\,\rm cont}\rangle/\langle F_{\lambda,\,\rm feat}\rangle$, following Section~5.3 of \citet{2024ApJ...976...82T}. To retain robust measurements, we require $\sigma(I_{\rm mol})\leq0.75$. Figure~\ref{fig:mol_index_teff} shows the resulting indices as a function of $T_{\mathrm{eff}}$, with the observed sources colored by their fitted $[\mathrm{M/H}]$.

The H$_2$O and CH$_4$ indices strengthen systematically toward lower $T_{\mathrm{eff}}$, consistent with the behavior established by near-infrared classification studies \citep{McLean2003NIRSPEC,Geballe2002Tclass,Burgasser2006Tclass}. Both indices show a change in slope, with a more rapid increase between approximately 1300 and 1100~K. This coordinated behavior is consistent with rapid chemical evolution across the L/T regime, where CH$_4$ becomes increasingly favored and the contrast of the water bands grows as the atmosphere cools \citep{SaumonMarley2008,Marley2021Sonora}. The residual dispersion likely reflects secondary dependencies on gravity and metallicity \citep{Allers2013UCD}, together with a small number of outliers caused by fluctuations in individual SPHEREx measurements within the index bandpasses.

The CO and CO$_2$ indices show a coherent but nonmonotonic temperature dependence. At $T_{\mathrm{eff}}\gtrsim1500$~K, neither index exhibits a strong overall temperature trend. Below approximately 1500~K, both indices increase, reach maxima near 1000~K, and then decline toward lower temperatures, with CO$_2$ showing the larger overall amplitude. Both indices also exhibit substantial dispersion at fixed $T_{\mathrm{eff}}$, particularly CO$_2$. These trends are broadly consistent with \citet{2026arXiv260700543R}. Using a curated L0--Y4 sequence and independently defined molecular indices, they find that CH$_4$ varies smoothly toward later and cooler spectral types, while the CO and CO$_2$ indices rise through the L/T sequence and then decline for cooler, CH$_4$-dominated objects, with a maximum near the mid-T regime.

The shared behavior of the CO and CO$_2$ indices suggests that they respond to related low-temperature atmospheric processes rather than to independent trends. Their increase below approximately 1500~K likely reflects changes in the thermal and chemical structure across the L/T regime that strengthen the CO and CO$_2$ bands relative to the local continuum \citep{2007ApJ...655.1079L,2012ApJ...756..172M,2011ApJ...734...73T,Sorahana2012AKARI}. Below approximately 1000~K, the decline of both indices is plausibly associated with the increasing partitioning of carbon into CH$_4$, which reduces the conditions favorable for strong CO and CO$_2$ absorption in the observable photosphere \citep{2014ApJ...797...41Z}. Some individual objects retain relatively strong CO absorption below approximately 1000~K, most likely because vertical mixing quenches the CO abundance by transporting CO-rich gas from deeper and hotter atmospheric layers \citep{2011ApJ...738...72V}.

These empirical sequences therefore establish a clear dominant temperature progression for H$_2$O and CH$_4$, whereas the larger dispersion of CO and CO$_2$ indicates sensitivity to additional physical parameters and measurement scatter.

\subsection{Model Comparison and Atmospheric-Parameter Sensitivity}
\label{subsec:molecular_model_comparison}

The model sequences in Figure~\ref{fig:mol_index_teff} allow us to assess both the relative performance of the SAND and ATMO2020++ grids and the diagnostic value of the molecular indices. For H$_2$O and CH$_4$, the two model families predict broadly similar sequences and reproduce the main observed temperature dependence reasonably well. For CO and CO$_2$, however, the model predictions diverge substantially, with ATMO2020++ providing the closer overall match to the observed relations.

The solar-metallicity gravity sequences further show that changes in $\log g$ affect all four indices. The dispersion in CO and CO$_2$ should therefore not be interpreted as a purely metallicity-driven effect. To quantify the relative sensitivity to metallicity and gravity, we examined the ATMO2020++ grid at fixed $T_{\mathrm{eff}}$ after removing the dominant temperature dependence. Over $T_{\mathrm{eff}}=800$--1200~K, the CO$_2$ index is more sensitive per dex to $[\mathrm{M/H}]$ than to $\log g$: a 1~dex change in $\log g$ produces an index change comparable to that produced by only approximately 0.3--0.6~dex in $[\mathrm{M/H}]$.

We also examined combinations of the H$_2$O, CH$_4$, CO, and CO$_2$ indices within the ATMO2020++ grid after removing the dominant $T_{\mathrm{eff}}$ dependence. The test results show that combinations involving H$_2$O and/or CO$_2$ improve the sensitivity to $[\mathrm{M/H}]$, but none cleanly separates the joint $[\mathrm{M/H}]$--$\log g$ parameter space. We therefore conclude that the four-index space can reduce, but does not fully break, the metallicity--gravity degeneracy. We do not use the SAND grid to draw quantitative conclusions about this degeneracy because its predicted CO and CO$_2$ sequences are substantially offset from the observed relations.

The fitted low-metallicity sources are concentrated mainly at $T_{\mathrm{eff}}\sim1300$--2000~K, whereas warmer sources tend to lie closer to solar metallicity in Figure~\ref{fig:mol_index_teff}. This apparent concentration should be interpreted cautiously. The relevant temperature range is dominated by L dwarfs and approaches the L/T transition, where the indices remain largely CO dominated and CH$_4$ becomes prominent only at lower temperatures. In this regime, the metal-poor SAND sequence predicts lower CO$_2$ indices than the solar and metal-rich sequences, while most observed CO$_2$ indices lie below the SAND predictions. Lowering $[\mathrm{M/H}]$ during spectral fitting can therefore move the models toward the observed CO$_2$ values and bias the inferred parameters toward lower metallicity. In addition, cloud and dust opacity can reshape the L-dwarf pseudo-continuum over the same wavelength range. The apparent low-metallicity concentration near the cloudy L-dwarf and L/T-transition regime may therefore reflect a combination of CO$_2$-driven fitting bias and cloud-related model mismatch rather than an intrinsic population trend.

Overall, the molecular indices trace both the dominant temperature sequence and secondary atmospheric dependencies across the ultracool-dwarf population. H$_2$O and CH$_4$ provide the clearest empirical temperature diagnostics, whereas CO and CO$_2$ carry additional information about metallicity and gravity.

\section{Conclusion}\label{sec:conclusion}

Our main results show how a homogeneous SPHEREx sample links observed ultracool dwarf spectrophotometry to atmospheric parameters, bolometric luminosities, evolutionary constraints, and molecular trends. The main conclusions are as follows.

\begin{enumerate}
\item We constructed the SPHEREx Ultracool Dwarf spectrophotometric Atlas (SUDA) from SPHEREx QR2 data and assembled a final sample of 1675 ultracool objects with reliable 0.75--5~$\mu$m spectrophotometry. This sample provides a homogeneous set of atmospheric parameters, including $T_{\mathrm{eff}}$, $\log g$, and [M/H], derived within a uniform framework using the SAND and ATMO2020++ model grids.

\item By combining the best fitting models with the observed SPHEREx spectrophotometry, we directly calculated $L_{\rm bol}$ for sources with parallaxes. For sources without parallax information, we used XGBoost to estimate $L_{\rm bol}$ from $T_{\mathrm{eff}}$, $\log g$, and [M/H]. We also used radii together with $T_{\mathrm{eff}}$ and C23 \citep{2023A&AChabrier} evolutionary models to derive masses, ages, and evolutionary $\log g$. The evolutionary $\log g$ values differ systematically from the atmospheric $\log g$ inferred from spectral fitting in the 1700--2500~K range, where the fitted atmospheric $\log g$ values are typically lower by a median offset of about 1.05~dex. This discrepancy is most consistent with cloud-driven model dependence in L-dwarf atmospheres and coupled degeneracies among cloud properties, $\log g$, metallicity, and possible non-equilibrium chemistry.

\item We constructed an empirical spectrophotometric atlas in the $T_{\mathrm{eff}}$ and $\log g$ diagram directly from the observed SPHEREx spectrophotometry. The measurements were binned using the $T_{\mathrm{eff}}$ fitted with atmospheric models and the $\log g$ inferred from evolutionary models, providing an observational reference sequence for ultracool dwarfs over 0.75--5~$\mu$m that links spectrophotometric morphology to the adopted physical parameter scale.

\item The molecular index analysis reveals a coherent atmospheric sequence across the sample. H$_2$O and CH$_4$ strengthen toward lower $T_{\mathrm{eff}}$, with a steeper rise between $\sim$1300 and 1100~K, and the SAND and ATMO2020++ model sequences reproduce these trends in broadly consistent ways. CO and CO$_2$ show a shared nonmonotonic evolution: both rise below $\sim$1500~K, reach a turnover near $\sim$1000~K, and decline at lower temperatures. For these carbon-bearing indices, the two model families differ much more strongly, with ATMO2020++ providing the closer overall match to the observed relation. The model grid indicates that both $[\mathrm{M/H}]$ and $\log g$ can affect the CO/CO$_2$ indices, but that CO$_2$ has greater sensitivity to metallicity than to gravity at $T_{\mathrm{eff}}\sim800$--1200~K. Combining multiple indices, particularly those involving H$_2$O and/or CO$_2$ can reduce, but do not fully break, the $[\mathrm{M/H}]$--$\log g$ degeneracy.
\end{enumerate}

Taken together, this work provides a homogeneous empirical basis for studying ultracool dwarfs across the 0.75--5~$\mu$m range. By combining a large spectrophotometric atlas, uniformly derived atmospheric parameters, bolometric luminosities, constraints from evolutionary models, and molecular index diagnostics within a single sample, SUDA links observed spectrophotometric morphology to atmospheric chemistry, substellar cooling, and the boundary between stars and brown dwarfs. It therefore provides an observational basis for future population studies, model calibration, and further investigations of the coolest and lowest mass sources with future infrared space telescopes.

\appendix
\onecolumngrid
\setcounter{figure}{0}
\setcounter{table}{0}
\makeatletter
\@addtoreset{figure}{section}
\@addtoreset{table}{section}
\makeatother
\renewcommand{\thefigure}{\thesection\arabic{figure}}
\renewcommand{\thetable}{\thesection\arabic{table}}
\section{Sample Quality Control}\label{appendix:sample_qc_cal}

This appendix summarizes the screening steps that define the final SUDA analysis sample.

\subsection{High S/N Source Screening}\label{app:high_sn_screening}

We applied quality and astrophysical cuts to isolate a reliable set of SPHEREx spectrophotometric measurements for the main analysis. First, we filtered individual measurements using SPHEREx data quality flags. Following the SPHEREx Explanatory Supplement\footnote{\url{https://irsa.ipac.caltech.edu/data/SPHEREx/docs/SPHEREx_Expsupp_QR.pdf}}, we retained only measurements with flag value $2^{21}$.

For each source, we then computed the median S/N over retained wavelength points and excluded sources with median S/N $<2$. We also applied astrophysical cuts using the UltracoolSheet to reduce systematics from unresolved systems and complex environments. Specifically, we removed unresolved binaries/multiples (\texttt{multiple\_unresolved\_in\_this\_table}), sources flagged as problematic (\texttt{flag}), and objects associated with active star forming regions (e.g., Taurus, Chamaeleon, Lupus, and IC~348) identified through \texttt{age\_category}.

After these cuts, we performed visual inspection and photometric consistency checks to mitigate contamination from blended sources within the SPHEREx beam. The retained SPHEREx spectrophotometry was compared against available 2MASS $J$, $H$, $K_s$ (or MKO $J$, $H$, $K$ when 2MASS was unavailable) and WISE $W1$ photometry. Synthetic broadband measurements were computed using filter profiles from the SVO Filter Profile Service\footnote{\url{http://svo2.cab.inta-csic.es/theory/fps3/}} and compared with catalog photometry as described below. The synthetic photometry calculation is identical to that used for the flux calibration analysis in Section~\ref{subsec:flux_cali}.

When SPHEREx fluxes were significantly higher than catalog photometry, we inspected WISE and ZTF images to assess contamination by nearby sources within $6$--$10^{\prime\prime}$ of the target; clearly contaminated objects were removed. When SPHEREx fluxes were significantly lower, we checked SPHEREx images and source alignment to identify possible positional offsets or centroid mismatches.

After all quality and selection cuts, the final sample contains 1675 ultracool stars and brown dwarfs with reliable SPHEREx spectrophotometry spanning $0.75$--$5~\mu$m.

\subsection{Blue-end Contamination Screening} \label{app:emissionline}

We identified a subset of measurements with anomalously elevated blue-end flux relative to the best fitting atmospheric models. The excess is most prominent in Band~1 (0.744--1.116\,$\mu$m) and can affect spectral classification, atmospheric fitting, and broadband synthetic photometry if left unmasked.

At least two mechanisms can produce this behavior. First, diffraction spikes or scattered light from a very bright primary star can affect a faint ultracool companion even when the companion lies outside the $6$--$10^{\prime\prime}$ close-neighbor range used for the blend checks above. HIP~73786B, for example, is separated from the bright primary HIP~73786A by about $68^{\prime\prime}$; because the system is only about 18~pc away, the primary is still bright enough on the sky to affect the blue end of the companion spectrophotometry. Second, atmospheric emission can elevate the same wavelength range, especially near the He~I line at 1.083\,$\mu$m. This atmospheric feature can remain in the extracted spectrophotometry, even after the standard quality-flag filtering, and therefore requires manual inspection. A detailed description of these atmospheric emission features is provided in the SPHEREx Explanatory Supplement.

\begin{figure}[htbp]
  \centering
  \includegraphics[width=0.8\textwidth]{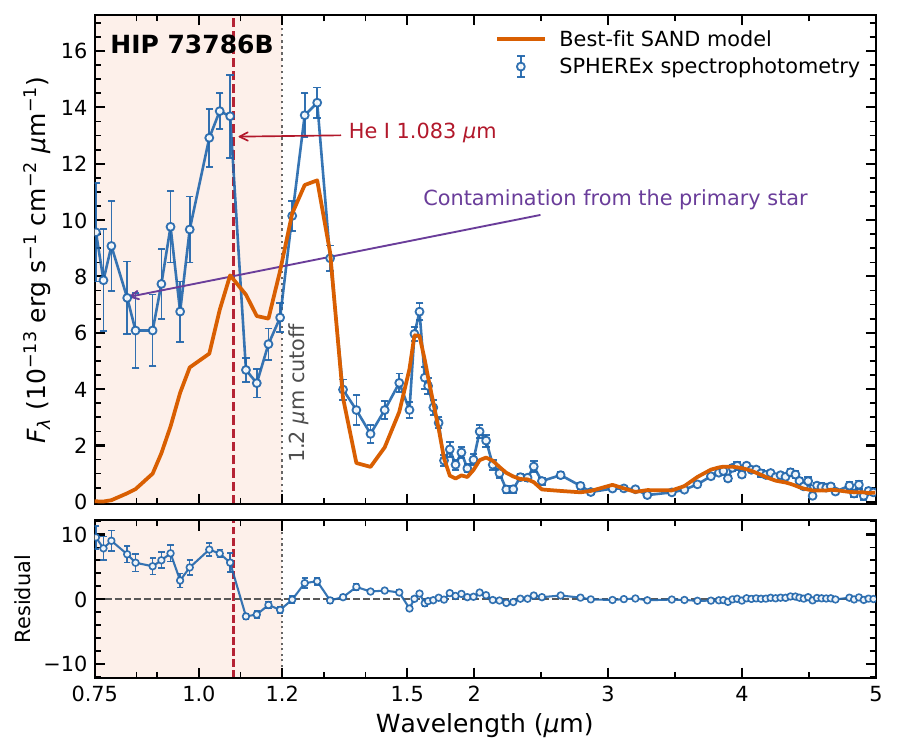}
  \caption{Example of blue-end contamination screening for HIP~73786B. The red dashed line marks the atmospheric He~I emission line at 1.083\,$\mu$m, and the dotted gray line marks the adopted 1.2\,$\mu$m cutoff. The shaded short-wavelength region is excluded from the subsequent analysis for affected sources. The bluest excess points are likely affected by contamination from the bright primary HIP~73786A. The lower panel shows the raw residuals, defined as $F_{\rm SPHEREx}-F_{\rm model}$, which highlight the blue excess associated with contamination by the earlier-type primary.}
  \label{fig:He_emission}
\end{figure}

Figure~\ref{fig:He_emission} illustrates the conditions for HIP~73786B, which was selected as an example because it shows both relevant effects: a broad blue-end excess likely associated with contamination from the bright primary HIP~73786A and an additional feature near the atmospheric He~I line at 1.083\,$\mu$m. The raw residuals, defined as $F_{\rm SPHEREx}-F_{\rm model}$, emphasize the blue excess and make contamination by the earlier-type primary more apparent. For affected sources, we performed source-by-source checks and excluded the contaminated blue-end measurements by truncating the SPHEREx spectrophotometry shortward of 1.2\,$\mu$m. Only the 1.2--5\,$\mu$m range was retained for subsequent analysis.

\section{Sample Catalog Table}\label{appendix:sample_catalog_table}
This appendix provides a representative portion of the SUDA catalog table, which contains the main quantities derived for all sources in this work.

\movetabledown=1.6in
\begin{rotatetable*}
\begin{deluxetable*}{lrrrrlrlllrrrrl}
  \setlength{\tabcolsep}{0.15cm}
  \tabletypesize{\scriptsize}
  \tablecaption{Summary table of information for all sources in the SUDA sample.}\label{tab:first30_trunc}
  \tablehead{
    \colhead{Name} & \colhead{RA} & \colhead{Dec} &
    \colhead{$T_{\rm eff}$} & \colhead{$\log g$} & \colhead{Model} &
    \colhead{Dist} & \colhead{SpT$_{\rm lit}$} & \colhead{SpT} &
    \colhead{Cut} & \colhead{$J_{\rm syn}$} & \colhead{$H_{\rm syn}$} &
    \colhead{$Ks_{\rm syn}$} & \colhead{$W1_{\rm syn}$} & \colhead{$\cdots$}
    \\
    \colhead{} & \colhead{(J2000)} & \colhead{(J2000)} &
    \colhead{(K)} & \colhead{} & \colhead{} &
    \colhead{(pc)} & \colhead{} & \colhead{} &
    \colhead{} & \colhead{(mJy)} & \colhead{(mJy)} &
    \colhead{(mJy)} & \colhead{(mJy)} & \colhead{}
  }
  \startdata
  SDSS J000013.54+255418.6 & 0.0565 & 25.9050 & $893_{-7}^{+7}$ & $4.77_{-0.22}^{+0.32}$ & ATMO & 14.1 & T4.5 & T4.0 & N & 1.610 & \ldots & \ldots & \ldots & \ldots \\
  PSO J000.2794+16.6237 & 0.2792 & 16.6241 & $2643_{-20}^{+19}$ & $4.09_{-0.18}^{+0.22}$ & SAND & 75.8 & M7.2 INT-G & M8.0beta & N & 1.709 & 2.083 & \ldots & \ldots & \ldots \\
  WISEA J000131.93-084126.9 & 0.3820 & -8.6900 & $2210_{-34}^{+37}$ & $4.42_{-0.28}^{+0.28}$ & SAND & 49.0 & L1 pec (blue) & \ldots & N & \ldots & 1.012 & \ldots & \ldots & \ldots \\
  LP 584-4 & 0.5259 & 1.2600 & $2577_{-27}^{+26}$ & $4.90_{-0.15}^{+0.07}$ & SAND & 20.7 & M9 & M8.0beta & N & \ldots & 25.249 & 23.552 & 13.379 & \ldots \\
  SDSS J000250.98+245413.8 & 0.7124 & 24.9039 & $1729_{-30}^{+29}$ & $4.77_{-0.14}^{+0.14}$ & SAND & -51.7 & L5.5 & \ldots & Y & 0.199 & 0.361 & 0.392 & 0.383 & \ldots \\
  2MASSI J0003422-282241 & 0.9263 & -28.3781 & $2635_{-12}^{+12}$ & $4.37_{-0.14}^{+0.13}$ & SAND & 40.4 & M7.5 & M8.0beta & N & 10.438 & 12.567 & 11.359 & 6.510 & \ldots \\
  SDSS J003524.44+144739.8 & 8.8518 & 14.7944 & $2371_{-53}^{+38}$ & $3.66_{-0.11}^{+0.23}$ & SAND & -120.9 & L0 & L0.0 & N & 0.345$^{\ast}$ & 0.409$^{\ast}$ & 0.449$^{\ast}$ & \ldots & \ldots \\
  WISE J004024.88+090054.8 & 10.1039 & 9.0152 & $913_{-23}^{+25}$ & $5.42_{-0.13}^{+0.06}$ & ATMO & 14.3 & T7 & T7.0 & N & 0.464$^{\ast}$ & \ldots & \ldots & 0.168$^{\ast}$ & \ldots \\
  2MASS J01311838+3801554 & 22.8267 & 38.0321 & $2029_{-20}^{+20}$ & $3.96_{-0.22}^{+0.25}$ & SAND & 24.4 & L4: & L1.0 & Y & 3.565$^{\ast}$ & 5.445$^{\ast}$ & \ldots & 3.957$^{\ast}$ & \ldots \\
  2MASS J01341675-0546530 & 23.5698 & -5.7814 & $1956_{-28}^{+20}$ & $4.29_{-0.18}^{+0.20}$ & SAND & -54.4 & L2 & \ldots & N & 0.433$^{\ast}$ & 0.705$^{\ast}$ & \ldots & \ldots & \ldots \\
  2MASS J01390900+8109597 & 24.7877 & 81.1667 & $2578_{-18}^{+19}$ & $3.82_{-0.21}^{+0.18}$ & SAND & 41.1 & L1 & M7.0gamma & N & \ldots & \ldots & 6.429$^{\ast}$ & 4.004$^{\ast}$ & \ldots \\
  WISEPA J020625.26+264023.6 & 31.6048 & 26.6730 & $1368_{-20}^{+33}$ & $3.69_{-0.13}^{+0.23}$ & SAND & 19.2 & L8 (red) & \ldots & Y & 0.295$^{\ast}$ & 0.514$^{\ast}$ & 0.666$^{\ast}$ & \ldots & \ldots \\
  \enddata
  \tablecomments{Only selected columns and representative rows are shown here. The full machine-readable catalog contains 1675 sources and 54 columns, including source names and coordinates (J2000); proper motions; SPHEREx-epoch propagated coordinates; atmospheric fitting results ($T_{\mathrm{eff}}$, $\log g$, [M/H], [$\alpha$/Fe], $\log(R^2/D^2)$, $\chi^2_r$, and best fitting model family); adopted distances; literature and SPLAT-matched spectral types; derived radii, masses, ages, and evolutionary $\log g$ values where available; $F_{\mathrm{bol}}$ and $\log(L_{\mathrm{bol}}/L_\odot)$; the short-wavelength fit-cut flag; and SPHEREx synthetic photometry in the $J$, $H$, $K$, and $W1$ bands with uncertainties. Cut: Y means that anomalous short-wavelength excess was identified and measurements shortward of 1.2~$\mu$m were excluded from this work, while N means that no such truncation was applied. SpT$_{\rm lit}$ is the literature spectral type, while SpT is the spectral type obtained in this work from SPLAT matching. Spectral types have a typical uncertainty of approximately $\pm1$ subtype, and entries with a colon have an uncertainty of approximately $\pm2$ subtypes. Blank entries in the SpT column indicate cases in which the corresponding SPHEREx spectrophotometry has insufficient S/N or incomplete blue wavelength coverage for reliable spectral type matching. Model denotes the atmospheric model that provides the best fit solution. The radius listed in the full electronic table is computed from $\log(R^2/D^2)=\log_{10}(R^2/D^2)$ and the adopted distance $D$. Some distance and radius entries are marked with negative signs as flags for quantities inferred indirectly from the machine-learning-based Lbol estimates; these no-parallax sources are not assigned C23 evolutionary $\log g$ values. The $J_{\rm syn}$, $H_{\rm syn}$, $K_{\rm syn}$, and $W1_{\rm syn}$ columns list the synthetic SPHEREx photometry before any flux calibration. A trailing asterisk marks a source for which source-level flux calibration was applied elsewhere in the analysis; for these starred entries, the listed synthetic photometry is the pre-calibration value and should not be treated as a reliable calibrated flux. The full table is available in machine-readable format from Zenodo (DOI: \url{https://doi.org/10.5281/zenodo.20659256}).}
\end{deluxetable*}
\end{rotatetable*}

\onecolumngrid
\section{Comprehensive Reference List for Sample Compilation}\label{appendix:comprehensive_reference_list}

This appendix lists the literature references that underlie the SUDA sample compilation. The reference list includes the source discovery references, parallax references, infrared spectral type references, and 2MASS/MKO $J$, $H$, $Ks/K$ and WISE $W1$ photometric references used for the objects in the SUDA catalog.

\begin{deluxetable}{l}[!ht]
  \tabletypesize{\scriptsize}
  \tablecaption{Literature References Associated with the SUDA Sample\label{tab:sample}}
  \tablehead{\colhead{References}}

  \startdata
  \parbox[t]{0.96\textwidth}{\citet{1959LowOBGiclas}; \citet{1979ApJLiebert}; \citet{1979nlcsLuyten}; \citet{1981MNRASReid}; \citet{1983ApJProbst}; \citet{1985MNRASGilmore}; \citet{1988MNRASHawkins}; \citet{1991AJBessell}; \citet{1991AJSchneider}; \citet{1991ApJSKirkpatrick}; \citet{1992ApJSLeggett}; \citet{1993AJTinney}; \citet{1993AJTinney2}; \citet{1993ApJKirkpatrick}; \citet{1993ApJTinney}; \citet{1994ApJSKirkpatrick}; \citet{1995AJKirkpatrick}; \citet{1995AJReid}; \citet{1996ApJSLeggett}; \citet{1996MNRASTinney}; \citet{1997AJKirkpatrick}; \citet{1997MNRASThackrah}; \citet{1997MNRASGizis}; \citet{1997PASPGizis}; \citet{1998AATinney}; \citet{1999AASDelfosse}; \citet{1999AJMarti}; \citet{Kirkpatrick1999L}; \citet{1999ApJStrauss}; \citet{1999ApJBurgasser}; \citet{2000AAScholz}; \citet{2000AJReid}; \citet{2000AJFan}; \citet{2000AJKirkpatrick}; \citet{2000AJGizis}; \citet{2000ApJBurgasser}; \citet{2000ApJLeggett}; \citet{2001AAJahreiss}; \citet{2001AAPhanBao}; \citet{2001AJWilson}; \citet{2001ApJSRuiz}; \citet{2002AALodieu}; \citet{2002AJSchneider}; \citet{2002AJReid}; \citet{2002AJCruz}; \citet{2002AJHawley}; \citet{2002AJReid2}; \citet{2002AJLe}; \citet{Burgasser2002T}; \citet{Geballe2002Tclass}; \citet{2002ApJGizis}; \citet{2002ApJLe}; \citet{2002MNRASScholz}; \citet{2002MNRASDobbie}; \citet{2003AAPokorny}; \citet{2003AAKendall}; \citet{2003AJLiebert}; \citet{2003AJBurgasser}; \citet{2003AJLe}; \citet{2003AJGizis}; \citet{2003AJTinney}; \citet{2003AJCruz}; \citet{2003AJReid}; \citet{2003AJBurgasser2}; \citet{2003AJReid2}; \citet{2003IAUSWilson}; \citet{2003tmc..book.....C}; \citet{2004AAHambaryan}; \citet{2004AAKendall}; \citet{2004AAReyle}; \citet{2004AAScholz}; \citet{2004AAScholz2}; \citet{2004AJBurgasser}; \citet{2004AJVrba}; \citet{2004AJKnapp}; \citet{2004ApJBurgasser}; \citet{2005AADeacon}; \citet{2005AALodieu}; \citet{2005AACrifo}; \citet{2005AJLe}; \citet{2005AJBochanski}; \citet{2005AJTinney}; \citet{2005AJEllis}; \citet{2005PASPReid}; \citet{2006AAPhanBao}; \citet{Skrutskie2006TwoMASS}; \citet{2006AJChiu}; \citet{2006AJRiaz}; \citet{Burgasser2006Tclass}; \citet{2006ApJBurgasser2}; \citet{2006MNRASPhanBao}; \citet{2006PASPLiebert}; \citet{2007AAKendall}; \citet{2007AADeacon}; \citet{2007AALeeuwen}; \citet{2007AJCruz}; \citet{2007AJReid}; \citet{2007AJLooper}; \citet{2007ApJBurgasser}; \citet{2007ApJArtigau}; \citet{2007ApJSLuhman}; \citet{2007MNRASKendall}; \citet{2007MNRASFolkes}; \citet{2008AADelorme}; \citet{2008AAScholz}; \citet{2008AJWest}; \citet{2008AJReid}; \citet{2008ApJBurgasser}; \citet{2008ApJMetchev}; \citet{2008ApJBurgasser2}; \citet{2008ApJLooper}; \citet{2008ApJKirkpatrick}; \citet{2008ApJBurgasser3}; \citet{2008MNRASPhanBao}; \citet{2008MNRASLodieu}; \citet{2008MNRASDayJones}; \citet{2009AASchilbach}; \citet{2009AAScholz}; \citet{2009AAZhang}; \citet{2009AJFaherty}; \citet{2009AJSheppard}; \citet{2009ApJSivarani}; \citet{2009ApJBurgasser}; \citet{2009ApJRadigan}; \citet{2009MNRASDeacon}; \citet{2010AAScholz}; \citet{2010AAScholz2}; \citet{2010AAMarti}; \citet{2010AJSchmidt}; \citet{2010AJSchmidt2}; \citet{2010AJBurgasser}; \citet{2010ApJBowler}; \citet{2010ApJBurgasser}; \citet{2010ApJArtigau}; \citet{2010ApJSKirkpatrick}; \citet{2010MNRASZhang}; \citet{2010MNRASBurningham}; \citet{2010MNRASGoldman}; \citet{2010MNRASBurningham2}; \citet{2011AAScholz}; \citet{2011AJWest}; \citet{2011AJAlbert}; \citet{2011AJDeacon}; \citet{2011AJGizis}; \citet{2011ANPhanBao}; \citet{2011ApJShkolnik}; \citet{2011ApJGeiss}; \citet{2011ApJLiu}; \citet{Cushing2011Y}; \citet{2011ApJSKirkpatrick}; \citet{2011MNRASMurray}; \citet{2011PhDTLooper}; \citet{2012AALodieu}; \citet{2012AALodieu2}; \citet{2012AJDeshpande}; \citet{2012ApJFaherty}; \citet{2012ApJLuhman}; \citet{2012ApJSDupuy}; \citet{2012MNRASFolkes}; \citet{2012yCat.2314....0L}; \citet{2013AABihain}; \citet{2013AAManjavacas}; \citet{2013AJMarocco}; \citet{Allers2013UCD}; \citet{2013ApJCastro}; \citet{2013ApJBest}; \citet{2013ApJLiu}; \citet{2013ApJSMace}; \citet{2013MNRASDayJones}; \citet{2013MNRASLodieu}; \citet{2013MNRASBurningham}; \citet{2013MNRASSmart}; \citet{2013MNRAS.435.2474L}; \citet{2013Msngr.154...35M}; \citet{2013PASPThompson}; \citet{2014AJNewton}; \citet{2014AJSchneider}; \citet{2014AJCushing}; \citet{2014AJMann}; \citet{2014AJAberasturi}; \citet{2014ApJLuhman}; \citet{2014ApJKirkpatrick}; \citet{2014ApJLuhman2}; \citet{2014ApJDeacon}; \citet{2014ApJGagliuffi}; \citet{2014ApJTinney}; \citet{2014MNRASDawson}; \citet{2014MNRASSmith}; \citet{2014MNRASLodieu}; \citet{2014yCat.2328....0C}; \citet{2015AJBurgasser}; \citet{2015AJSchmidt}; \citet{2015AJKellogg}; \citet{2015ApJ...799...37L}; \citet{2015ApJBaron}; \citet{2015ApJBest}; \citet{2015ApJSGagne}; \citet{2015ApJSTerrien}; \citet{2015MNRASMarocco}; \citet{2015MNRASCardoso}; \citet{2015MNRASBeami}; \citet{2016AASkrzypek}; \citet{2016AJWeinberger}; \citet{2016ApJSchneider}; \citet{2016ApJAller}; \citet{2016ApJRobert}; \citet{2016ApJLiu}; \citet{2016ApJSKirkpatrick}; \citet{2016ApJSFaherty}; \citet{2016NaturGillon}; \citet{2016PhDTAller}; \citet{2016yCat.2343....0E}; \citet{2017AJSchneider}; \citet{2017AJShkolnik}; \citet{2017AJKellogg}; \citet{2017AJDahn}; \citet{2017ApJBest}; \citet{2017ApJSGagne}; \citet{2017MNRASZhang}; \citet{2018AAReyle}; \citet{2018AAPe}; \citet{2018ApJGagne}; \citet{2018ApJSTinney}; \citet{2018MNRASSmart}; \citet{2019AATorres}; \citet{2019AJKiman}; \citet{2019AJGreco}; \citet{2019ApJMatsuoka}; \citet{2019ApJGagliuffi}; \citet{2019ApJSKirkpatrick}; \citet{2019MNRASZhang}; \citet{2019MNRASZhang2}; \citet{2020AAScholz}; \citet{2020AJBest}; \citet{2020ApJFaherty}; \citet{2020ApJZhang}; \citet{2020MNRASMarocco}; \citet{2021AJ....161...42B}; \citet{2021ApJSKirkpatrick}; \citet{Marocco2021CatWISE2020}; \citet{2021yCat.2367....0M}; \citet{2022AJSchneider}; \citet{2022ApJVos}; \citet{GaiaDR3}; \citet{2023AJSchneider}; \citet{2023ApJSanghi}; \citet{2024arXivBest}}
  \enddata
  \tablecomments{This table lists the literature references associated with the SUDA sample, including the discovery references, parallax references, infrared spectral type references, and 2MASS/MKO $JHK$ and WISE $W1$ photometric references for all sources in the catalog.}
\end{deluxetable}

\clearpage
\section{Machine Learning Details for $\log L_{\mathrm{bol}}$ Estimation}\label{appendix:ml_lbol_details}

This appendix summarizes the XGBoost method used to estimate $L_{\mathrm{bol}}$ for sources without parallaxes.

\subsection{Machine Learning Method and Validation}
\label{app:ml_method}

We train an XGBoost regressor to predict $\log(L_{\mathrm{bol}}/L_\odot)$ from $T_{\mathrm{eff}}$, $\log g$, and $[\mathrm{M/H}]$. The model is trained with a squared error objective and evaluated using the mean absolute error (MAE) in $\log(L_{\mathrm{bol}}/L_\odot)$. The regression target is
$y=\log(L_{\mathrm{bol}}/L_\odot)$,
and the feature vector is
${x}=(T_{\mathrm{eff}},\,\log g,\,[\mathrm{M/H}])$.
These features are chosen because they are available for essentially all objects and are physically linked to luminosity through the relation between radius and temperature.

The training set is constructed from sources with reliable luminosities based on parallax. To mitigate contamination by problematic entries (e.g., unphysical solutions driven by erroneous photometry or distances), we apply additional quality cuts based on the inferred radius. Specifically, we retain only sources with radii in the range $0.7$--$2.5\,R_{\mathrm{Jup}}$. Radii are computed using the flux scaling parameter $\log(R^2/D^2)$ obtained from the nested sampling analysis in combination with literature parallaxes. After these selections, the final training set contains 1046 sources. The trained model is then applied to 550 sources without parallax measurements, which constitute the prediction sample of this study.

The training labels have heterogeneous uncertainties, which we incorporate through sample weighting. Each training object is assigned a weight
\begin{equation}
  w = \frac{1}{\sigma_{\log L}^2 + \sigma_0^2},
\end{equation}
where $\sigma_{\log L}$ is the reported uncertainty in $\log(L_{\mathrm{bol}}/L_\odot)$. The constant $\sigma_0$ acts as a regularization floor, preventing a small number of measurements with unrealistically small reported luminosity uncertainties (often driven by very small parallax errors) from dominating the optimization. We set $\sigma_0$ to the 20th percentile of the $\sigma_{\log L}$ distribution ($\sigma_0 = 0.0248$ dex in our training sample).

Hyperparameters are selected using ten fold cross validation on the training set. We use the out of fold (OOF) MAE as the primary metric, since it summarizes typical prediction error for unseen objects drawn from the same distribution. The final model achieves an OOF MAE of $0.123$ dex. We optimize six hyperparameters and keep all others at XGBoost defaults. The adopted values are $\gamma=0$, learning rate $=0.02$, maximum tree depth $=3$, maximum number of leaves $=32$, number of estimators $=300$, and $\lambda=0$ (L2 regularization).

For sources without parallax, we propagate uncertainties in the input atmospheric parameters into the predicted luminosities by Monte Carlo sampling. For each object, we draw 3000 realizations of $(T_{\mathrm{eff}},\,\log g,\,[\mathrm{M/H}])$. The trained XGBoost model is evaluated for each realization, yielding a distribution of predicted $y=\log(L_{\mathrm{bol}}/L_\odot)$. We adopt the mean of the Monte Carlo prediction distribution as the predicted luminosity and use its standard deviation as the corresponding uncertainty.

\begin{figure*}[h]
  \centering
  \includegraphics[width=0.49\textwidth]{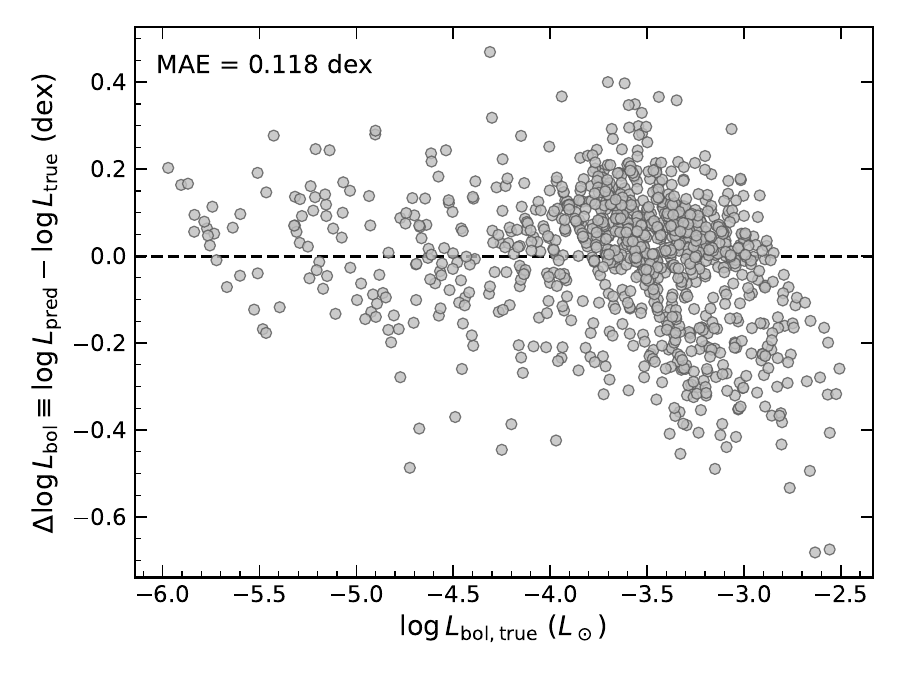}\hfill
  \includegraphics[width=0.49\textwidth]{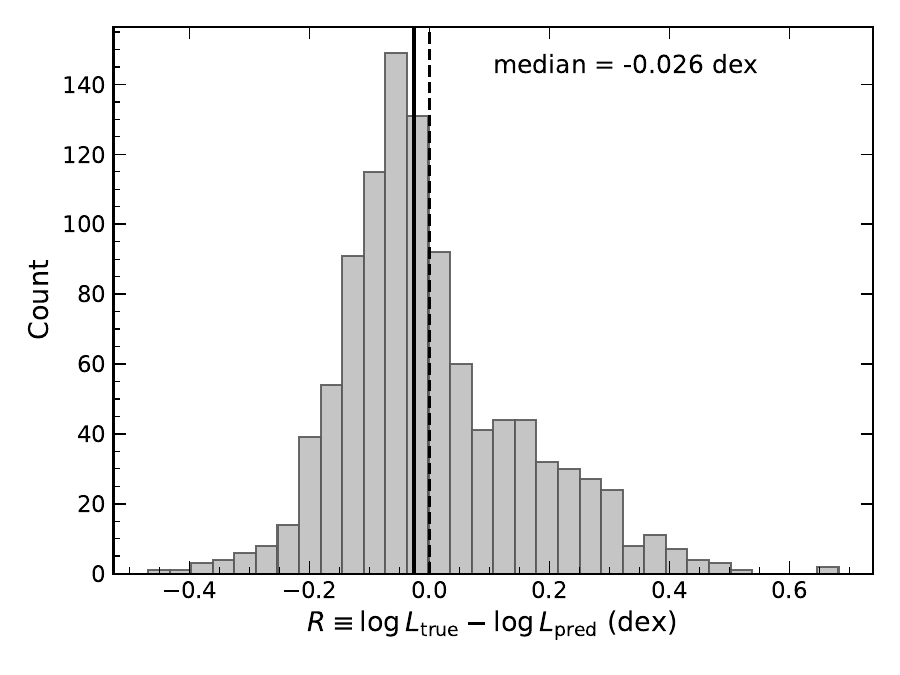}
  \caption{Left: prediction performance of the ML model for $\log L_{\mathrm{bol}}$ using the training samples. Right: residual distribution for $\Delta\log L_{\mathrm{bol}}$ in the training set.}
  \label{fig:ml_train_eval}
\end{figure*}

\subsection{Training Results Evaluation}

The trained XGBoost model reproduces the training labels with good overall accuracy, while showing mild bias at the highest luminosities. The final refit predictions for the training set have an MAE of 0.118 dex, and the out of fold MAE is 0.123 dex. Figure~\ref{fig:ml_train_eval} summarizes the training performance: the left panel shows predicted versus true $\log L_{\mathrm{bol}}$, and the right panel shows the residual distribution $\Delta \log L_{\mathrm{bol}}$.

At the high $L_{\mathrm{bol}}$ end, the left panel shows a systematic underestimation, which shifts the center of the residual distribution in the right panel toward negative values. This bias likely arises from two effects: (1) regression to the mean in a regularized model, which pulls predictions toward the bulk of the training set and suppresses the brightest tail; and (2) a biased distribution of training set $L_{\rm bol}$ in $(T_{\rm eff}, \log g)$ space, where the luminous objects occupy sparsely sampled regions near the coverage boundary, leaving the model weakly constrained there.

\section*{Data Availability}

The SUDA data products are publicly available from Zenodo at DOI: \url{https://doi.org/10.5281/zenodo.20659256}. The release includes the catalog, the 102-channel SPHEREx spectrophotometry, the preliminary SPHEREx bandpass table used for the approximate response-function treatments, the empirical spectrophotometric atlas products, and the trained XGBoost luminosity model. The JWST data presented in this article were obtained from the Mikulski Archive for Space Telescopes (MAST) at the Space Telescope Science Institute. The specific observations analyzed can be accessed via \dataset[doi: 10.17909/hryb-x508]{https://doi.org/10.17909/hryb-x508}.

\begin{acknowledgments}
  We thank the anonymous referee for the constructive comments and suggestions, which helped improve the quality of this paper. This work is supported by the National Natural Science Foundation of China (NSFC) through the projects 12588202, 12373028, 12322306, 12173047, and 12133002. S. W. and X. C. acknowledge support from the Youth Innovation Promotion Association of the CAS with Nos. 2023065 and 2023055. Z. T. thanks Furen Deng for helpful discussion. This work makes use of data products from the Spectro-Photometer for the History of the Universe, Epoch of Reionization and Ices Explorer (SPHEREx), which is a joint project of the Jet Propulsion Laboratory and the California Institute of Technology, and is funded by the National Aeronautics and Space Administration, and is hosted by IPAC as part of the IRSA archive (SPHEREx
  Quick Release Spectral Images, QR2 \citealt{2025IRSA652}). This work has benefited from the UltracoolSheet, maintained by Will Best, Trent Dupuy, Michael Liu, Aniket Sanghi, Rob Siverd, and Zhoujian Zhang, and developed from compilations by \citet{2012ApJSDupuy,2013Sci...341.1492D,2014ApJDeacon,2016ApJLiu,2018ApJS..234....1B,2021AJ....161...42B,2023ApJSanghi,2023AJSchneider}.
  This work is based on observations made with the NASA/ESA/CSA James Webb Space Telescope. The data were obtained from the Mikulski Archive for Space Telescopes (MAST) at the Space Telescope Science Institute.
\end{acknowledgments}






\software{astropy \citep{2013A&A...558A..33A,2018AJ....156..123A,2022ApJ...935..167A}, NumPy \citep{numpy}, SciPy \citep{scipy}, Matplotlib \citep{matplotlib}, UltraNest \citep{ultranest}, SPLAT \citep{2017ASInC..14....7B}, XGBoost \citep{xgboost}.
}

\bibliographystyle{aasjournalv7}
\bibliography{main}{}

\end{document}